\documentclass[preprint,5p]{elsarticle}
\usepackage[cmex10]{amsmath}
\usepackage{amssymb}
\usepackage{algorithm}
\usepackage[hyphens]{url}
\usepackage[noend]{algpseudocode}

\usepackage{multirow}
\usepackage[utf8]{inputenc}

\newcommand{\speci}[2][c]{\begin{tabular}[#1]{@{}l@{}}#2\end{tabular}}
\journal{Computer Physics Communications}

\begin{document}

\begin{frontmatter}

\title{The density matrix renormalization group algorithm on kilo-processor architectures: implementation and trade-offs}

\author[itk]{Csaba Nemes \corref{cor1}}
\ead{nemes.csaba@itk.ppke.hu}
\author[wigner]{Gergely Barcza}
\author[itk,sztaki]{Zoltán Nagy}
\author[wigner]{Örs Legeza}
\author[itk,sztaki]{Péter Szolgay}

\cortext[cor1]{Corresponding author}

\address[itk]{Faculty of Information Technology, Péter Pázmány Catholic University, Budapest, Hungary}

\address[sztaki]{Cellular Sensory and Wave Computing Laboratory, Computer and Research Automation Institute, Hungarian Academy of Sciences, Budapest, Hungary}
 
\address[wigner]{ Strongly Correlated Systems "Lend\"ulet" Research Group, Department of Theoretical Solid State Physics, Wigner Research Centre for Physics, Hungarian Academy of Sciences, Budapest, Hungary}

\begin{abstract}
In the numerical analysis of strongly correlated quantum lattice models one of the leading algorithms developed to balance the size of the effective Hilbert space and the accuracy of the simulation is the density matrix renormalization group (DMRG) algorithm, in which the run-time is dominated by the iterative diagonalization of the Hamilton operator.
As the most time-dominant step of the diagonalization can be expressed as a list of dense matrix operations, the DMRG is an appealing candidate to fully utilize the computing power residing in novel kilo-processor architectures.

In the paper a smart hybrid CPU-GPU implementation is presented, which exploits the power of both CPU and GPU and tolerates problems exceeding the GPU memory size.
Furthermore, a new CUDA kernel has been designed for asymmetric matrix-vector multiplication to accelerate the rest of the diagonalization.
Besides the evaluation of the GPU implementation, the practical limits of an FPGA implementation are also discussed.
\end{abstract}

\begin{keyword}
strongly correlated systems \sep DMRG \sep GPU acceleration \sep FPGA acceleration
\end{keyword}

\end{frontmatter}

\section{Introduction and Related Works}
DMRG is a variational numerical approach developed to treat low-dimensional interacting many-body quantum systems efficiently~\cite{White1992,Noack2004,Schollwock2005}.
In fact, it has become an exceptionally successful method to study the low energy physics of strongly correlated quantum systems which exhibit chain-like entanglement structure~\cite{Legeza2005}.
For example, it can be applied to simulate properties of anisotropic materials, such as polymers~\cite{Barford2005}, or to describe accurately the electronic structure of open $d$ shell molecules~\cite{Barcza2011}, which is beyond the capability of standard quantum chemical approaches.
Additionally, the interacting system of atoms trapped in an optical lattice, proposed as physical implementation of quantum computer,  is also tractable via DMRG~\cite{Lewenstein2007}.

The original DMRG algorithm~\cite{White1992} was introduced in 1992 by Steven R. White and was formulated as a single threaded algorithm. 
In the past various works have been carried out to accelerate the DMRG algorithm on shared  \cite{Hager2004}~\cite{Alvarez2012} and distributed memory~\cite{Chan2004algorithm,Yanai2009,Yamada2011,Rincon2010} architectures, however, none of them took advantage of recent kilo-processor architectures: graphical processing unit (GPU) and field-programmable gate array (FPGA).

One of the first parallelizations was~\cite{Hager2004} converting the projection operation to matrix-matrix multiplications and accelerating them via OpenMP interface.
In~\cite{Yamada2011} a similar approach was presented for distributed memory environment (up-to 1024 cores) optimizing the communication between the cores, while in~\cite{Rincon2010} the acceleration of the computation of correlation function had been investigated.
Recently,~\cite{Alvarez2012} presented an acceleration on shared memory architectures exploiting SU(2) symmetries, while \cite{White2013} proposed a novel direction for paralellization via a modification of the original serial DMRG algorithm.

Graphical processing unit has been successfully employed in neighboring research areas to accelerate matrix operations.
In~\cite{Yu201155} GPU is used to accelerate tensor contractions in plaquette renormalization states (PRS), which can be regarded as an alternative technique to tensor product states (TPS) or the DMRG algorithm. 
In~\cite{Cawkwell2012} the second-order spectral projection (SP2) algorithm has been accelerated, which is an alternative technique to calculate the density matrix via a recursive series of generalized matrix-matrix multiplications.

In this paper we present the first attempt (to our best knowledge) to investigate how the DMRG method can utilize the enormous computing capabilities of novel kilo-processor architectures (GPU, FPGA). 
In case of GPU a smart hybrid CPU-GPU acceleration is presented, which tolerates problems exceeding the GPU memory size, consequently, supporting wide range of problems and GPU configurations.
Contrary to the previous acceleration attempts not only the projection operation is accelerated, but further parts of the diagonalization are also computed on the GPU.
In case of FPGA the performance limits of a possible implementation are estimated and discussed.

The rest of the paper is organized as follows. 
Section~\ref{SecModels} describes the models which are used as test cases to demonstrate the operation of the algorithm.
Symmetries which can be exploited to decrease the computational requirements of the algorithm and the algorithm itself are presented in Sections~\ref{SecSymm} and~\ref{SecAlg}, respectively.
Acceleration on GPU is presented in three sections (\ref{SecRuntime},~\ref{SecAccelMatrix} and~\ref{SecAccelProjection}), while limits of an FPGA implementation are described in Section~\ref{SecFPGA}.
Finally, implementation results and conclusions are given in Sections~\ref{SecImp} and~\ref{SecCon}, respectively.

\section{Investigated models}\label{SecModels}
In order to illustrate the underlying features of the algorithm it is applied to the so-called spin-$\frac{1}{2}$ Heisenberg model and the spin-$\frac{1}{2}$ Hubbard model.
The selected models describe how to compute the Hamiltonian of the system of interest, while the main task is to find some of the low-lying eigenvalues and eigenvectors of the Hamiltonian by a diagonalization algorithm. 
In practice instead of solving  the problem for the complete Hilbert space directly, various  physical phenomena can be exploited to reduce the complexity of the problem.

\subsection{Heisenberg model}
The Heisenberg model describes the physics of magnetic systems and provides theoretical description of various experimental measurements.
In the model a magnetic system is simulated on a lattice of interacting \textit{spins}.
A microscopic magnetic moment (spin) is localized at each lattice site $j$ and described by a quantized, two-valued variable, $\sigma_j \in \{ \uparrow,\downarrow \}$, related to the two possible orientations of the spin.
Limiting the interactions to only neighboring spins -- which is often a good approximation --  
the Hamiltonian of the model is written as
\begin{equation}
    H = \frac{1}{2} \sum_{j=1}^{N-1} \left( S^+_{j} S^-_{j+1} + S^-_{j} S^+_{j+1}\right) 
     + \Delta  \sum_{j=1}^{N-1} S^z_j S^z_{j+1} 
     \label{ham}
\end{equation}
where $S_j^+, S_j^-$ operators change, while $S_j^z$ measures the orientation of the spin on lattice site $j$.
The overall behavior of the system can be tuned via the relevant parameter $\Delta$.
The explicit matrix representation of an operator ${\cal O}_j$ acting on site $j$ of a chain with $N$ spins is given as
\begin{equation}
{\cal O}_{j}=\bigotimes_{i=1}^{j-1} \mathbb{I} \otimes {\cal O}  \otimes \bigotimes_{i=j+1}^{N} \mathbb{I}
\label{op}
\end{equation}
where $\mathbb{I}$ is the identity and $\cal O$ is one of the followings
\begin{equation}
    S^+ = \begin{pmatrix} 0&1\\ 0&0 \end{pmatrix},~
    S^- = \begin{pmatrix} 0&0\\ 1&0 \end{pmatrix},~
    S^z = \frac{1}{2}\begin{pmatrix} 1&0\\ 0&-1 \end{pmatrix}.
\end{equation}
The Hamiltonian of $N$ spins acts on the tensor product space of dimension $2^N$, that is the dimension of the complete Hilbert space grows exponentially as the size of the system increases.

\subsection{Hubbard model}
The Hubbard model was introduced to describe electrons in solids to characterize the transition between insulating and conducting systems.
The single-band Hubbard model provides appropriate description of low temperature systems where all particles are in the lowest Bloch band and the long-ranged interactions between the particles can be neglected due to strong screening effects~\cite{Gebhard97}.
More recently various multi-band Hubbard models have been applied to high-temperature superconductivity~\cite{Anderson97}
and systems of higher spin to understand the behavior of optically trapped ultracold atoms~\cite{Lewenstein2007}.

In the general spin-$F$ system each lattice site is characterized by $2F+1$ two dimensional vectors.
Each vector is assigned with a distinct label (from $\{-F,-F+1,...,F-1,F\}$) called spin polarization value (denoted by $\sigma$). 
A vector assigned to a spin polarization $\sigma$ describes two orthogonal states: the site is occupied ($[0;1]$) by the particle of spin polarization $\sigma$ or not ($[1;0]$).
As a consequence, a lattice site of spin-$F$ possesses $2^{2F+1}$ internal degrees of freedom.

The lattice model of interacting particles of spin-$F$ consists of two competing terms: the kinetic term, which describes the tunneling of particles between neighboring lattice sites, and the local potential term, which describes on-site density-density interaction measuring the attraction or repulsion between the interacting particles.
The single-band, fermionic Hubbard model of spin-$F$ is defined on a chain with $N$ sites as 
\begin{equation}
    H = -t \sum_{j=1}^{N-1}\sum_{\sigma=-F}^F \big(c^{\dagger}_{j,\sigma} c^{\phantom\dagger}_{j+1,\sigma}+\mathrm{h.c.} \big) 
    + \frac{U}{2} \sum_{j=1}^N \sum_{\sigma\neq \sigma'} n^{\phantom\dagger}_{j,\sigma} n^{\phantom\dagger}_{j,\sigma'}
    \label{ham2}
\end{equation}
where $t$ measures the hopping amplitude between neighboring sites and $U$ is the interaction strength. 
Creation and annihilation operator acting on site $j$ with spin polarization $\sigma$, denoted as $c^\dagger_{j,\sigma}$ and $c_{j,\sigma}$, adds or removes a particle located on site $j$ with spin polarization $\sigma$.
The particle density of spin polarization $\sigma$ on site $j$ is measured by operator $n^{\phantom\dagger}_{j,\sigma}= c_{j,\sigma}^\dagger c_{j,\sigma}^{\phantom\dagger}$. 
The explicit matrix representation of an operator ${\cal O}_{j,\sigma}$ acting on site $j$ and polarization $\sigma$ is constructed as
\begin{equation}
{\cal O}_{j,\sigma} =\bigotimes_{i=1 }^{F'(j-1)} \Phi \otimes {\cal O}_\sigma \otimes \bigotimes_{i=F'(j+1)}^{F'N} \mathbb{I}
\label{op2}
\end{equation}
\begin{equation}
{\cal O}_\sigma =
\bigotimes_{i= -F}^{\sigma-1}  \Phi \otimes {\cal O} \otimes \bigotimes_{i=\sigma+1}^{F}\mathbb{I}
\label{fermi_op2}
\end{equation}
\begin{equation}
	\Phi = \begin{pmatrix} 1&0\\ 0&-1 \end{pmatrix}
\end{equation}
where $F'=2F+1$, $\mathbb{I}$ is the identity, $\Phi$ is the fermionic phase-factor and ${\cal O}$ is one of the followings
\begin{equation}
    c^\dagger = \begin{pmatrix} 0&0\\ 1&0 \end{pmatrix},~
    c = \begin{pmatrix} 0&1\\ 0&0 \end{pmatrix}.
\end{equation}
The Hamiltonian describing the spin-$F$ system of $N$ lattice sites acts on the tensor product space of dimension $2^{F'N}$, and similarly to the Heisenberg model, the dimension of the complete Hilbert space blows up exponentially.
Comparing to the bosonic operators of the Heisenberg model, the key differences in the construction of operators are the appearance of internal quantum number, $\sigma$, and the presence of the phase-factor describing the antisymmetric nature of fermionic systems. To ease the comparison of the two models only the $F=\frac{1}{2}$ case is presented, however, the observed tendencies are valid for higher F values.

\section{Symmetries to be exploited}\label{SecSymm}

In many systems the Hamilton operator does not change the value of a measurable quantity, i.e., it commutes with the operator connected to that measurable quantity.
These operators are called symmetry operators and can be used to cast the Hilbert space to smaller independent subspaces\cite{Toth2008}.
Consequently, instead of solving a large matrix eigenvalue problem, the eigenvalue spectrum can be determined by solving several smaller problems.
In the Heisenberg model the total spin projection, $S_z=\sum_{j=1}^N S^z_j$, is such a symmetry operator.
Meanwhile, in the Hubbard model of spin-$F$ the total particle number associated to each spin polarization $\sigma$, $N_{\sigma}=\sum_{j=1}^N n_{j,\sigma}$, is conserved.
Thus, the distinct quantum numbers helps to partition the Hilbert space into multiple independent subspaces corresponding to a given combination of quantum number values.

\begin{figure}[b]
  \begin{center}
    \includegraphics[height=5cm]{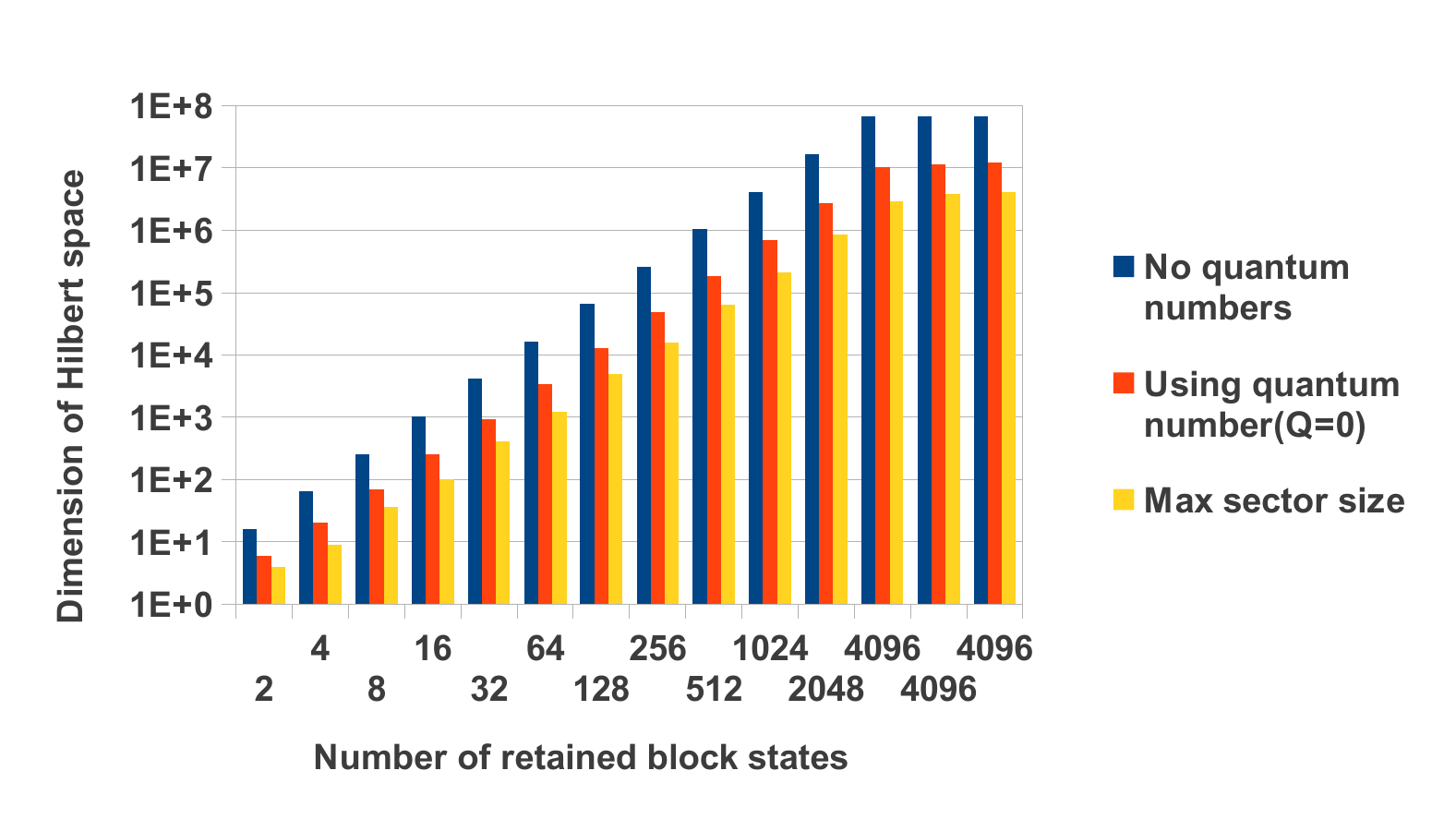}
  \end{center}
  \caption{Exploiting the projection symmetry in the Heisenberg model the Hilbert space of the superblock can be restricted to the subspace corresponding to Q=0.}
  \label{FigUsingQnum}  
\end{figure}

A given symmetry operator shares the same eigenvectors of the Hamiltonian, thus the eigenstates of the Hamiltonian can be labeled by the eigenvalues of the symmetry operator (\textit{quantum number}, $Q$), and the Hilbert space can be decomposed into subspaces (\textit{sectors}) spanned by the eigenvectors of each quantum number value~\cite{Cornwell1997}. Introducing a quantum number based representation, the sparse operators (Eqs.~\ref{op},~\ref{op2}) can be decomposed to a set of smaller but dense matrices, furthermore the Hamiltonian operator (Eqs.~\ref{ham},~\ref{ham2}) becomes blockdiagonal.

\section{Algorithm}\label{SecAlg}

The DMRG approach has two phases, in the \textit{infinite-lattice algorithm} the approximated Hilbert space of a finite system of $N$ interacting spins is built up iteratively, while in the optional \textit{finite-lattice algorithm} the number of the interacting spins is fixed and further iterations are carried out to increase the accuracy of the computed results.
As in both cases the iterations are very similar, for the sake of simplicity, we consider only the infinite-lattice algorithm.
The detailed description of the algorithm can be found in the original work~\cite{White1992} and various reviews~\cite{Noack2004,Schollwock2005}, here only the key steps of an iteration of the infinite-lattice algorithm are summarized in Algorithm~\ref{alg1} providing the basis of our analysis.

\begin{algorithm}
\caption{One iteration of the infinite-lattice algorithm}
\label{alg1}
\begin{algorithmic}[1]
\State Load a left and a right block.
\State Form the superblock configuration.
\State Compute the lowest eigenstate of the superblock Hamilton $H_{SB}$. (Davidson method)
\For {each block}
\State Construct the density matrix for the given block from the lowest eigenstate.
\State Compute the eigenvalues of the density matrix. (Lanczos method)
\State Renormalize the basis of the block by keeping states with high eigenvalues.
\EndFor
\end{algorithmic}
\end{algorithm}

\begin{figure}[t]
  \begin{center}
    \includegraphics[height=5cm]{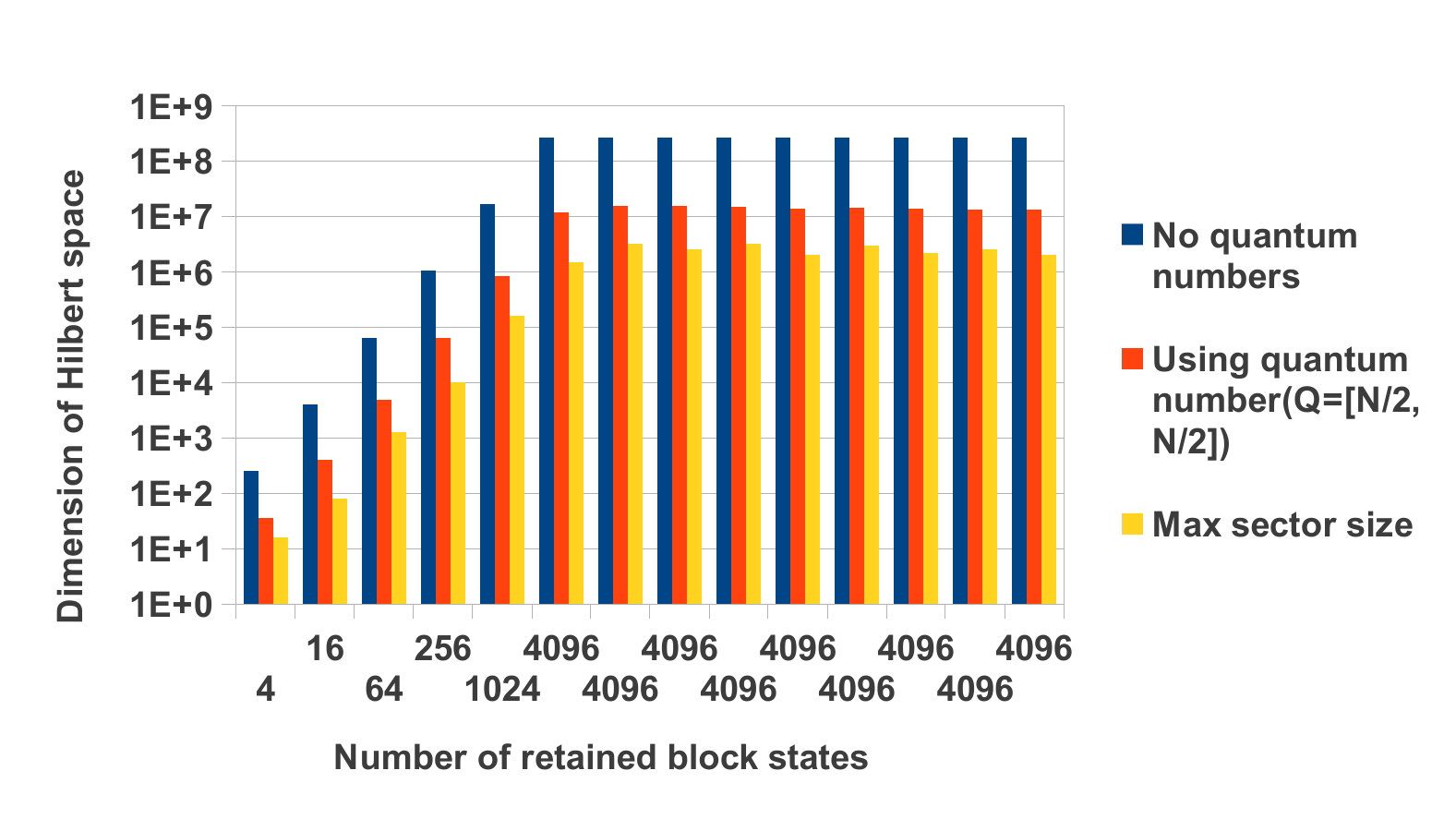}
  \end{center}
  \caption{Exploiting the conservation of particle number in the spin-$\frac{1}{2}$ Hubbard model the Hilbert space of the superblock can be restricted to the subspace corresponding to Q=[$\frac{N}{2}$,$\frac{N}{2}$].}
  \label{FigUsingQnumHub}  
\end{figure}

In the two-site DMRG procedure four subsystems (left block describing $l$ sites, $1$ site, $1$ site, right block describing $r$ sites) compose the finite system of $N=(l+2+r)$ sites called \textit{superblock}. 
The sites contained in each block are described maximally by $m$, optimally chosen states, which can be significantly smaller than the exactly required $q^l$ or $q^r$ basis, where $q$ is the degree of freedom of one site.
As the central sites of the superblock are represented exactly by q-q states, the size of the superblock Hilbert space is $q^2m^2$. 
Considering, however, the symmetries mentioned above, the problem can be restricted to a subspace of the superblock corresponding to a particular $Q$ value. 
E.g., in case of Heisenberg and Hubbard models the size of the superblock Hilbert space can be reduced significantly as demonstrated in Figures~\ref{FigUsingQnum} and~\ref{FigUsingQnumHub}, respectively.
It is, however, clear that even using symmetry operators the dimension of the reduced space grows exponentially with the size of the lattice (if no truncation is done).

The infinite-lattice algorithm starts with the four site configuration, where each block contains a single spin.
In each iteration step both blocks are enlarged by a single site, making the complete system increase by two, until the desired system size, $N$, is reached. 
In each iteration of the DMRG algorithm, the lowest-lying eigenvector of the corresponding superblock Hamiltonian ($H_{SB}$) is obtained by the iterative Davidson or Lanczos algorithm.
(In the paper the Davidson algorithm is considered.)
From the lowest eigenstate the density matrix is constructed which carry the information how to optimally truncate the basis of the enlarged block ($m\ll q^{l+1}$) in order to keep the problem size manageable~\cite{Legeza2003}.

The most time-consuming part of a full iteration is the step of the Davidson routine which carries out the projection operation ($X'=H_{SB}X$).
Instead of constructing and storing the enormous $H_{SB}$ matrix of size ${\cal O}\left(m^4\right)$ explicitly, it is computationally favorable to obtain the projected vector $X'$ directly from the matrices of size ${\cal O}\left(m^2\right)$ composing $H_{SB}$.

The $H_{SB}$ can be directly expressed by the operators of the original four subsystems (\textit{l}-1-1-\textit{r strategy)} or by the operators of two intermediate systems (\textit{LR strategy}), so called \textit{enlarged blocks}, which come from the contraction of each block with its neighboring site.
In the current implementation only the second strategy is investigated, however, the first one is also straightforward and will be included in the near future.

There are several practical benefits of these strategies. 
First of all, the memory bandwidth limited matrix-vector multiplication (BLAS Level 2) is converted to matrix-matrix multiplication (BLAS Level 3) which can be efficiently accelerated. 
Secondly, skipping of the explicit Kronecker multiplications not only restructures the computation, but decreases the number of operations.
Finally, both strategies drastically decrease the size of the matrices which take part in the operations and thus the memory footprint of the algorithm. 
In case of LR and \textit{l}-1-1-\textit{r} strategy the largest matrix has a size of $O((mq)^2)$ and $O((m)^2)$, respectively.
The second strategy is more favorable in extreme situations when the GPU memory is limited and $q$ (internal degrees of freedom) is large (e.g. spin-$F$ Hubbard model with large $F$).

\subsection{\textit{LR strategy}}

In the \textit{LR strategy} the $H_{SB}$ is expressed with operators $A_{\alpha}^{(L)}$ and $B_{\alpha}^{(R)}$ defined on the left ($L:=l+1$) and right ($R:=r+1$) enlarged blocks, respectively, as
\begin{equation}
H_{SB} = \sum_{\alpha} A_{\alpha}^{(L)}\otimes B_{\alpha}^{(R)}~,
\end{equation}
where the index $\alpha$ iterates over the distinct operator combinations required to construct the superblock Hamiltonian. 
Exploiting Kronecker multiplication properties, the projected vector $X'$ can be computed by matrix-matrix multiplications as
\begin{equation}
\tilde{X}'= \sum\limits_{\alpha} A_{\alpha}^{(L)} \tilde{X} B_{\alpha}^{(R)T} ~,
\label{EqProjection}
\end{equation}
where vector $X$ of size $[B^{col}A^{col}]$ is reshaped to matrix $\tilde{X}$ of size $[B^{col},A^{col}]$ and vector $X'$ of size $[B^{row}A^{row}]$ is reshaped to matrix $\tilde{X}'$ of size $[B^{row},A^{row}]$.

\begin{algorithm}[h!]
\caption{The computation of the projected vector $X'$ in case of \textit{l}-1-1-\textit{r strategy} strategy.}
\label{algL11r}
\small
\begin{algorithmic}[1]
\Require{$size(X)=[D^{col} C^{col} B^{col} A^{col}]$}
\Function{projectX\_l11r}{$A,B,C,D,X$}
	\State $X_1$=reshape($X$) \textbf{as} size($X_1$)=$[D^{col}, C^{col}, B^{col}, A^{col}]$
	\For {each $(i,j)$} $X_1'(:,:,i,j)$=$DX_1(:,:,i,j)$
	\EndFor		
	\For {each $(i,j)$} $X_1''(:,:,i,j)$=$X_1'(:,:,i,j)C^T$
	\EndFor
	\State $X_2$=reshape($X_1''$) \textbf{as} size($X_2$)=$[D^{row} C^{row}, B^{col}, A^{col}]$
	\For {each $(i)$} $X_2'(:,:,i)$=$X_2(:,:,i)B^T$
	\EndFor
	\State $X_3$=reshape($X_2'$) \textbf{as} size($X_3$)=$[D^{row} C^{row} B^{row}, A^{col}]$
	\State $X_3'$=$X_3 A^T$
	\State $X'$=reshape($X_3'$) \textbf{as} size($X'$)=$[D^{row} C^{row} B^{row} A^{row}]$
	\State \Return $X'$
\EndFunction
\end{algorithmic}
\end{algorithm}

In the practical implementation Equation~\ref{EqProjection} operates on even smaller matrices as the operators are decomposed according to quantum numbers.
Instead of a sparse matrix $A^{(L)}$ several dense matrices $A_{q_i \rightarrow q_j}^{(L)}$ are stored representing how $A^{(L)}$ transforms the subspace (sector) corresponding to $q_i$ to the one corresponding to $q_j$.
To compute $X'$ in case of a given $A_{\alpha}^{(L)},B_{\alpha}^{(R)}$ operator pair all possible $A_{\alpha,q_i \rightarrow q_j}^{(L)}$ , $B_{\alpha,q_k \rightarrow q_l}^{(R)}$ transition pairs shall be submitted to
Equation~\ref{EqProjection} and each time only the corresponding $ik$ and $jl$ segments of $X$ and $X'$ shall be used as
\begin{equation}
\tilde{X}_{jl}'+= A_{\alpha,i \rightarrow j}^{(L)} \tilde{X}_{ik} B_{\alpha,k \rightarrow l}^{(R)T}
\label{EgProjectionDecomposed}
\end{equation}
where $\tilde{X}_{ik}$ and $\tilde{X}_{jl}'$ indicate the reshaped $ik$ and $jl$ segment of vector  $X$ and $X'$, respectively. Fortunately, the reshape operation has no computational cost as the data in the memory is untouched and only the row/col sizes are changing.

\subsection{\textit{l}-1-1-\textit{r strategy}}

In the \textit{l}-1-1-\textit{r strategy} the $H_{SB}$ is expressed by the operators of the four subsystems:
\begin{equation}
H_{SB} = \sum_{\alpha} A_{\alpha}^{(l)}\otimes a_{\alpha} \otimes b_{\alpha} \otimes B_{\alpha}^{(r)}~,
\end{equation}
where the index $\alpha$ again iterates over the distinct operator combinations required to construct the superblock Hamiltonian.

Similarly to the LR strategy, Kronecker multiplication properties can be exploited to compute the projected vector  $X'$ efficiently with matrix-matrix operations, however, in this case a more complicated data storage and several tensor multiplications are needed to avoid unnecessary memcopy operations. Using the procedure \textsc{projectX\_l11r()}, which computes the projected vector $X'$ for one matrix quadruplet and is described in Algorithm~\ref{algL11r}, the $H_{SB}$ can be calculated as
\begin{equation}
X'= \sum\limits_{\alpha} \textsc{projectX\_l11r} (A_{\alpha}^{(l)},a_{\alpha}^{\phantom\dagger},  b_{\alpha} , B_{\alpha}^{(r)},X) ~.
\label{EqProjectionlqqr}
\end{equation}

In the similar manner as shown in $LR$ strategy, $A^{(l)},a,b$ and $B^{(r)}$ operators can be decomposed according to quantum numbers and instead of large sparse matrix operations several smaller dense matrix operations shall be submitted to Algorithm~\ref{algL11r}. Furthermore, none of the reshape operations of Algorithm~\ref{algL11r} involves practical data movement, only the size descriptor variables are changing.
\begin{figure}[t]
  \begin{center}
    \includegraphics[height=5cm]{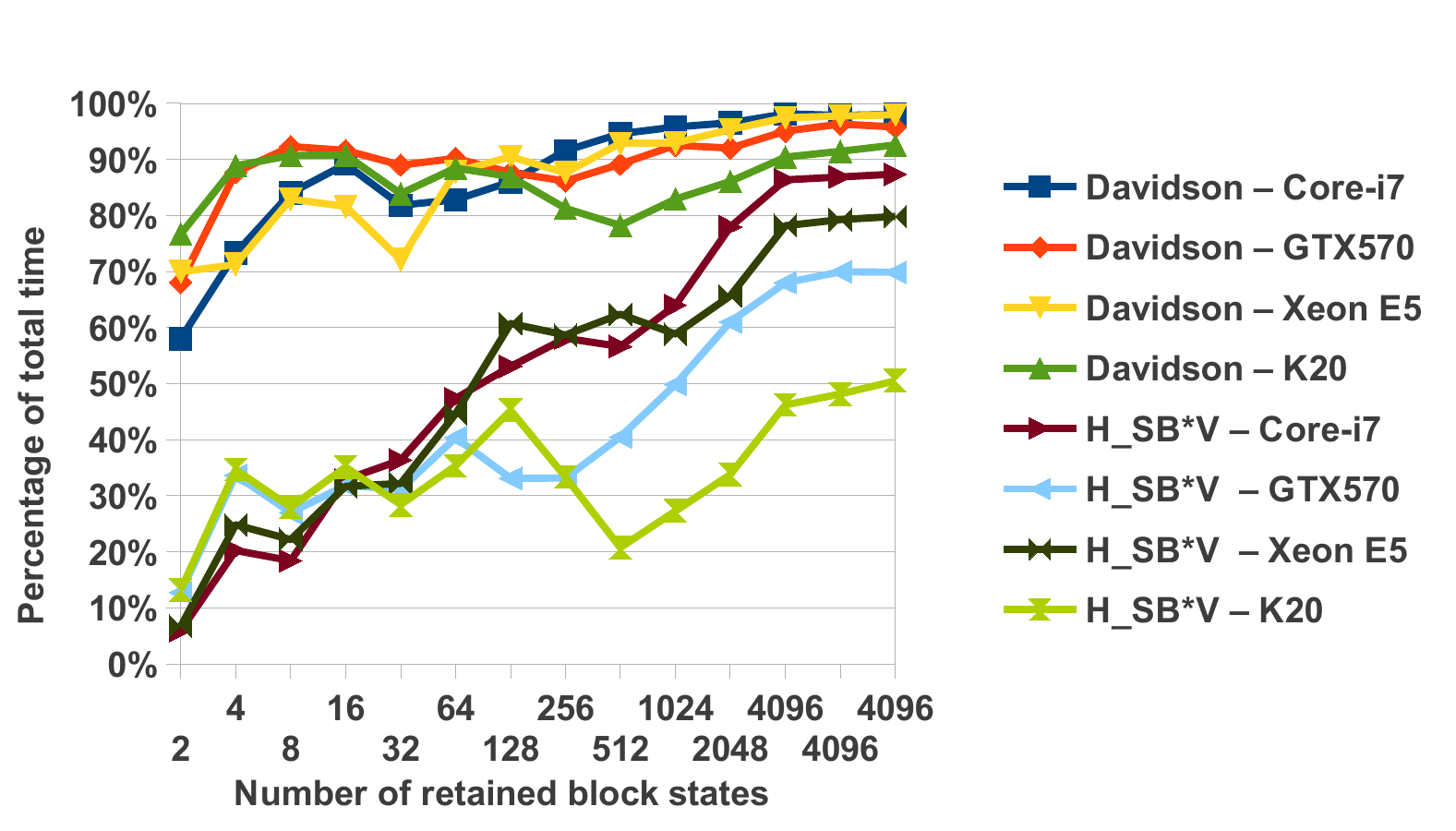}
  \end{center}
  \caption{Heisenberg model: Runtime of the Davidson algorithm and its $H_{SB}V$ operation compared to the total time of a DMRG iteration step as the number of retained block states increases. CPU-only versions are indicated by Core-i7 and Xeon E5, while hybrid versions are indicated by GTX 570 and K20.}
  \label{FigPercentage}
\end{figure}

\begin{figure}[h!]
  \begin{center}
    \includegraphics[height=5cm]{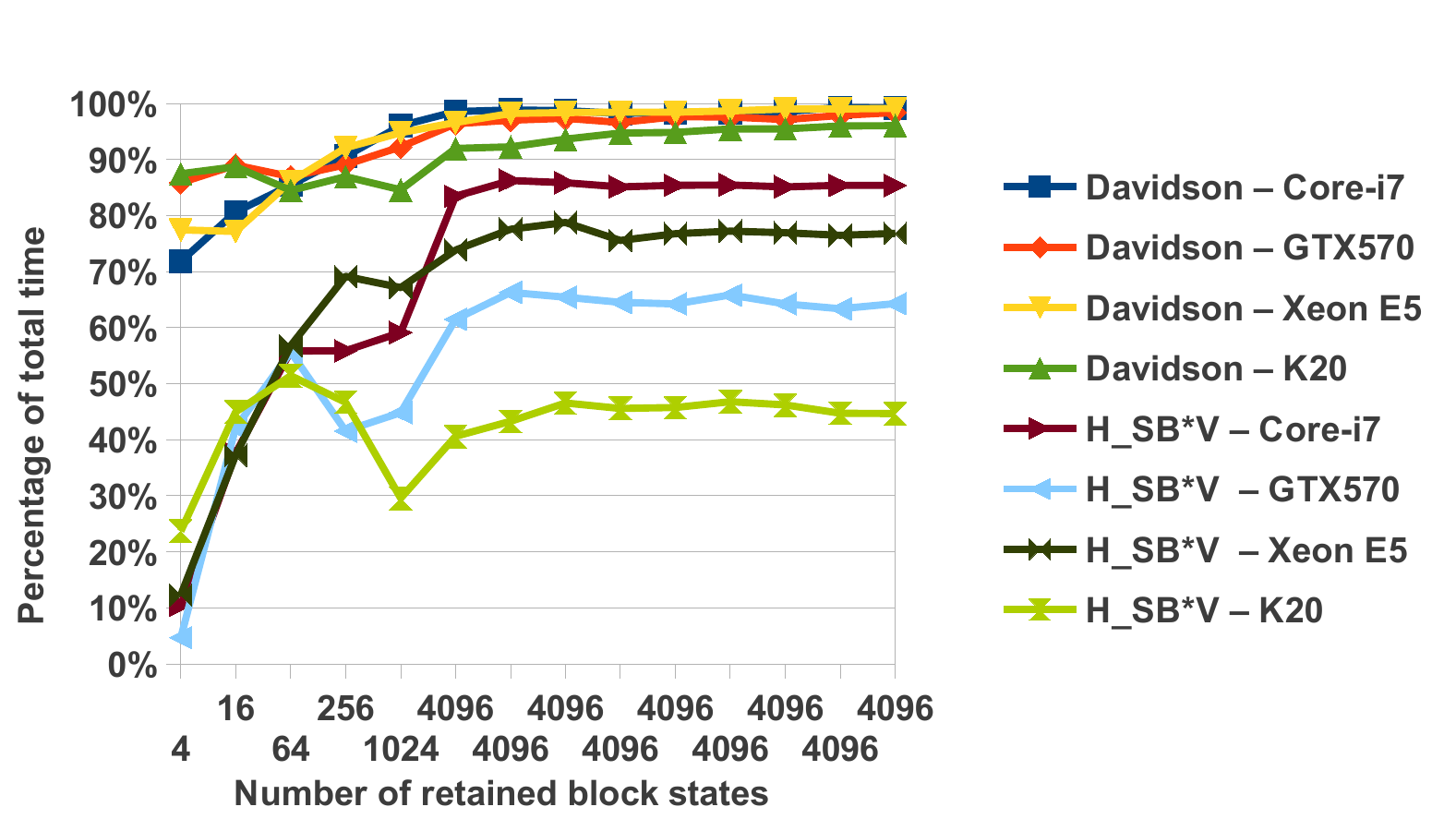}
  \end{center}
	\caption{Similar to Figure~\ref{FigPercentage} but for the Hubbard model.}
  \label{FigPercentageHub}
\end{figure}

\section{Runtime \& Parallelism}\label{SecRuntime}

In case of the Heisenberg and Hubbard model the runtime analysis of the DMRG algorithm is shown in Figures~\ref{FigPercentage} and~\ref{FigPercentageHub}, respectively.
As the Davidson algorithm, which is summarized in Algorithm~\ref{algDav}, is the most time-dominant part and takes more than 97\% of the total time in the CPU-only reference implementation, it has been chosen for acceleration.
Unfortunately, the full Davidson algorithm cannot be implemented on the GPU as the problem size in real world simulations usually exceeds the GPU memory size.
Instead, a hybrid approach shall be implemented, which can adjust the GPU workload according to the available GPU memory and the CPU-GPU performance ratio.

\begin{algorithm}[h!]
\caption{One iteration of the Davidson algorithm}
\label{algDav}
\begin{algorithmic}[1]
\Require{Previous $(i-1)$ basis vectors already computed.}
\Function{davidsonIter}{$i$}
    \State $W(:,i)=H_{SB}\cdot V(:,i)$ \Comment{BLAS-3: dgemm() }
	\State $B(:,i)=V^T \cdot W(:,i)$ \Comment{BLAS-2: dgemv\_trans() }
 	\State $[\lambda,y]\longleftarrow \text{get smallest eigvalue and vector of B}$ 
 	\State $x = V \cdot y$	\Comment{BLAS-2: dgemv() }
    \State $r =  - \lambda \cdot x + W \cdot y$
    \If{$\text{norm}(r)\approx 0$} 
	    \State return with $x$ and success
    \Else 
    	\State correct $r$
		\State // orthonormalize $r$ against $V$:
		\State $s=V^T \cdot r$\Comment{BLAS-2: dgemv\_trans() }
		\State $r=r-V \cdot s$\Comment{BLAS-2: dgemv() }
		\State normalize $r$ and append to $V$
		\State return without success
    	\EndIf
\EndFunction
\end{algorithmic}
\end{algorithm}

In the Davidson algorithm inherent parallelism can be observed at two levels.
First, at low level, all the matrix and vector operations can be accelerated.
Secondly, at the level of projection computation (line 2 in Algorithm~\ref{algDav}), which is the most time-dominant part of the Davidson algorithm itself taking more than 75\% of the total time, the projection can be computed as a sum of the independent $(AX)B^T$ operations (see Equation~\ref{EqProjection}).

At low level, the CPU part of the algorithm uses the Basic Linear Algebra Subroutine (BLAS) interface and the Intel MKL Library~\cite{IntelMKL} for algebraic operations including operator contractions, inner operations of both Davidson and Lanczos algorithms~\cite{Saad92} and operator transformations. 
Unfortunately, in the Davidson algorithm, all the operations except the projection are BLAS level 2 matrix-vector multiplications, which are bandwidth limited and not ideal for acceleration. There is block extension~\cite{Liu1978} of the algorithm, the so called Davidson-Liu, to determine a few of the lowest eigenvalues, where more than one candidate vectors are added at once resulting in BLAS level 3 operations, however, in the current DMRG implementation only the lowest eigenvalue is investigated.
The remaining option is to store as much data in GPU memory as possible and execute the corresponding operations on GPU.

At the level of projection operation the independence of matrix multiplications provides a straightforward hybrid parallelization and a future multi-GPU modification of the current implementation.
Acceleration can be improved by developing the appropriate scheduling of the matrix operations for different matrix sizes and architectures.

\begin{figure}[]
  \begin{center}
    \includegraphics[height=5cm]{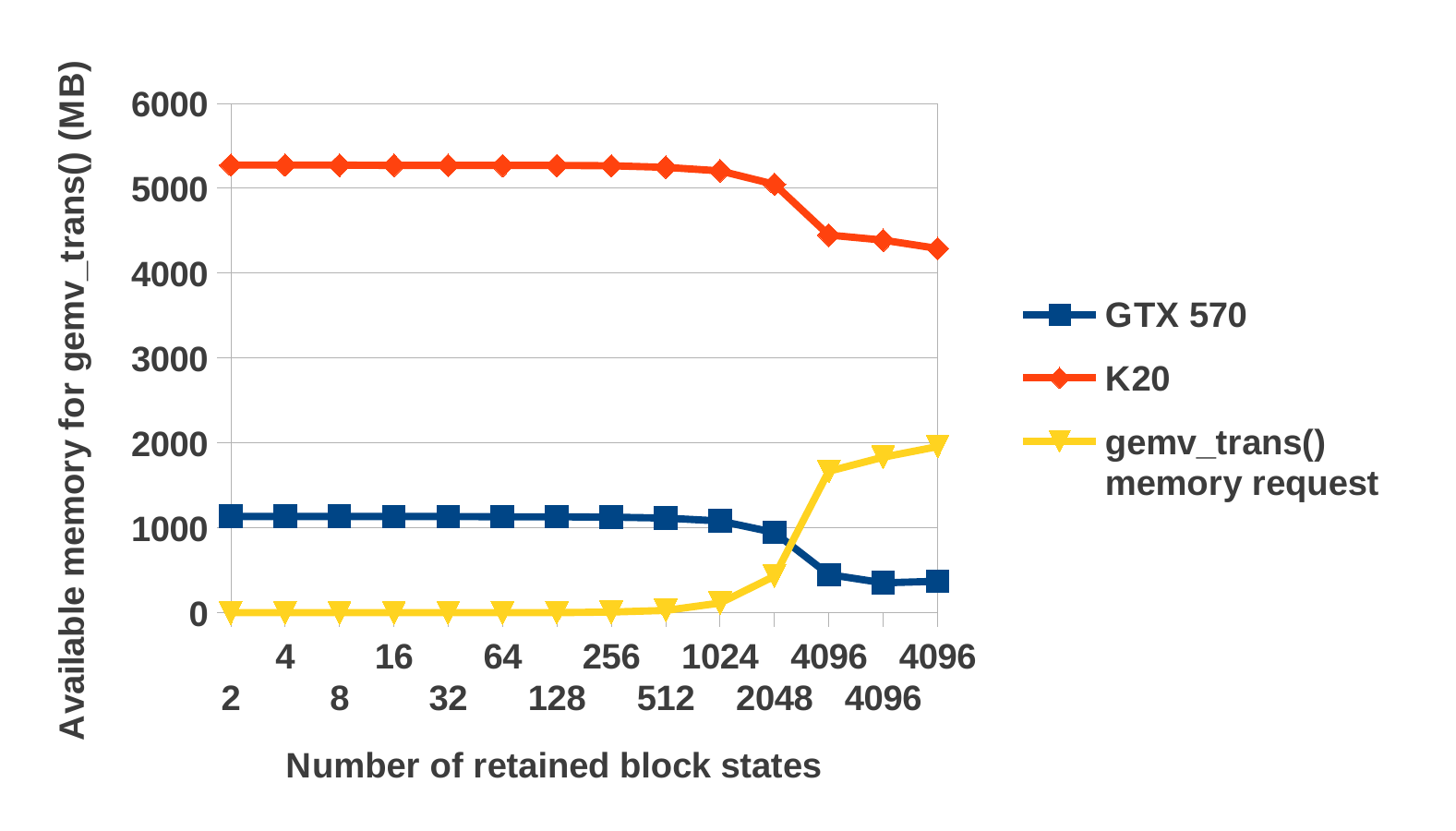}
  \end{center}
  \caption{Remained free GPU memory after the projection operation, which can be used for gemv\_trans() and the memory request of the gemv\_trans() in case of the Heisenberg model.}
  \label{FigGemvFootprint}
\end{figure}

\section{Accelerating matrix-vector multiplications}\label{SecAccelMatrix}

Jacobi-Davidson version~\cite{Sleijpen1996} of the original Davidson algorithm~\cite{Davidson1975} is a preconditioned subspace iteration technique~\cite{Saad92} aimed at computing a few of the extreme eigenpairs of large sparse symmetric matrices and commonly used in the DMRG implementations~\cite{Noack2004,Schollwock2005}.
In the presented work the~\cite{Sadkane1999} version of the algorithm (available in Netlib~\cite{Netlib}) is implemented.

In each iteration the subspace is extended with a new base vector ($V(:,i)$), which is stored in the memory accompanied by its projection ($W(:,i)$). 
As the size of these vectors can be very large (see Figures~\ref{FigUsingQnum},~\ref{FigUsingQnumHub}) depending on the model and the number of retained block states ($m$), they cannot be fully stored in the GPU memory.
However, in order to accelerate a matrix-vector multiplication with GPU, at least the matrix shall be stored in the GPU memory.
In the current implementation the matrix of the basis vectors ($V$), which is used four times (see comments in Algorithm~\ref{algDav}) in BLAS level 2 operations, has been selected to be stored, although the storage of the projected vector matrix ($W$) can be added later as well.

In each iteration the new basis vector is loaded to the GPU memory in the background (if there is available space) and the workload of the BLAS level 2 operations is shared between the CPU and GPU: CPU process the new basis vector, while GPU operates on the older ones.
With this technique the power of both CPU and GPU can be exploited and the transfer time of the matrix can be hidden.
The implementation is flexible: if there is no more space on the GPU or the CPU performance justifies it, more than one base vector can be processed on the CPU leaving less work for GPU.
As shown in Figure~\ref{FigGemvFootprint}, where the storage requirement of $V$ is compared with available free space after the projection operation, $V$ cannot be fully stored on a GTX 570 GPU even in case of the simple Heisenberg model ($m=4096$).

There are two types of BLAS level 2 operations: $V^TX$ and $VX$ indicated by \textit{gemv\_trans()} and \textit{gemv()} in Algorithm~\ref{algDav}, respectively.
In the first case the multiplier vector can be loaded while in the second case the result can be written in smaller parts (${\sim}5e5$) to overlap with the computation.

\subsection{gemv\_trans()}

\begin{figure}[t]
  \begin{center}
    \includegraphics[height=5cm]{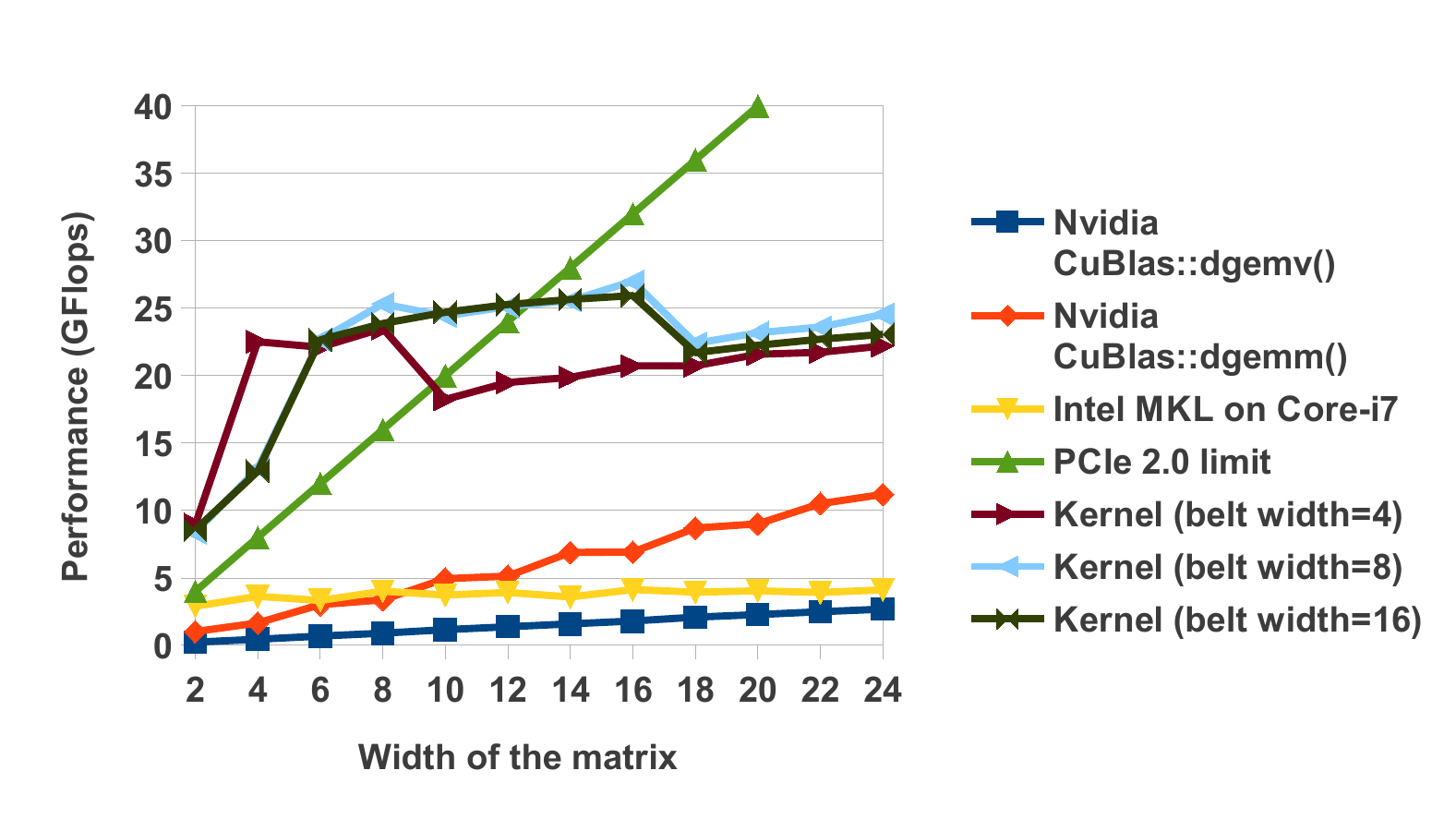}
  \end{center}
  \caption{GTX 570: Performance of the presented \textit{gemv\_trans()} kernel with different belt widths is compared to the performance of the available implementations in case of matrix height $5e5$. 
Additionally, the PCIe limit is displayed as PCIe throughput limits the performance of the GPU acceleration if the transfer of the multiplier vector cannot be avoided.}
  \label{FigGemvGtx5e5}
\end{figure}

\begin{figure}[t]
  \begin{center}
    \includegraphics[height=5cm]{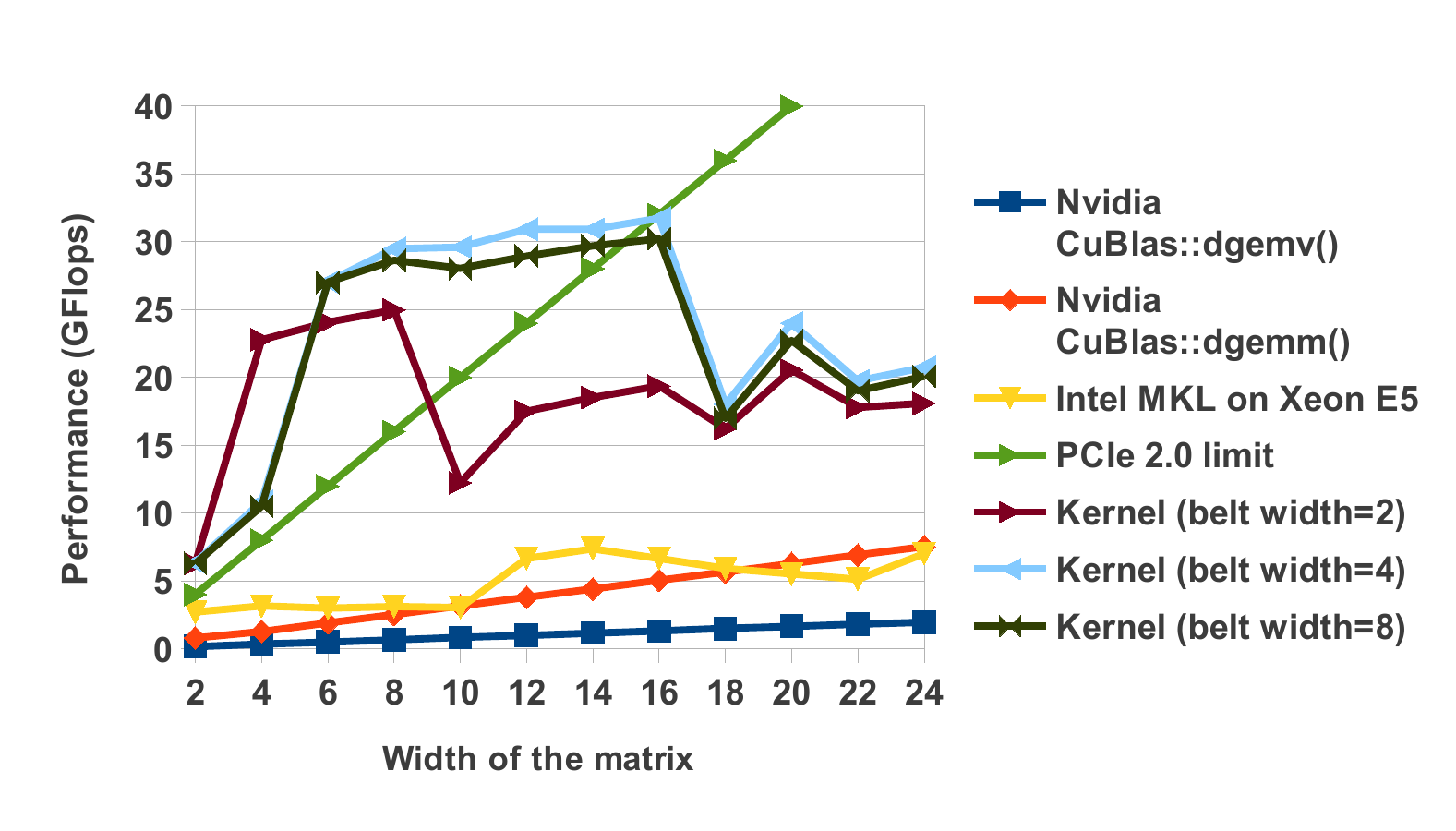}
  \end{center}
  \caption{K20, no shuffle operation in the kernel: Performance of the presented \textit{gemv\_trans()} kernel with different belt widths is compared to the performance of the available implementations in case of matrix height $5e5$. 
Additionally, the PCIe limit is displayed as PCIe throughput limits the performance of the GPU acceleration if the transfer of the multiplier vector cannot be avoided.}
  \label{FigGemvk20noshuffle}
\end{figure}

\begin{figure}[t]
  \begin{center}
    \includegraphics[height=5cm]{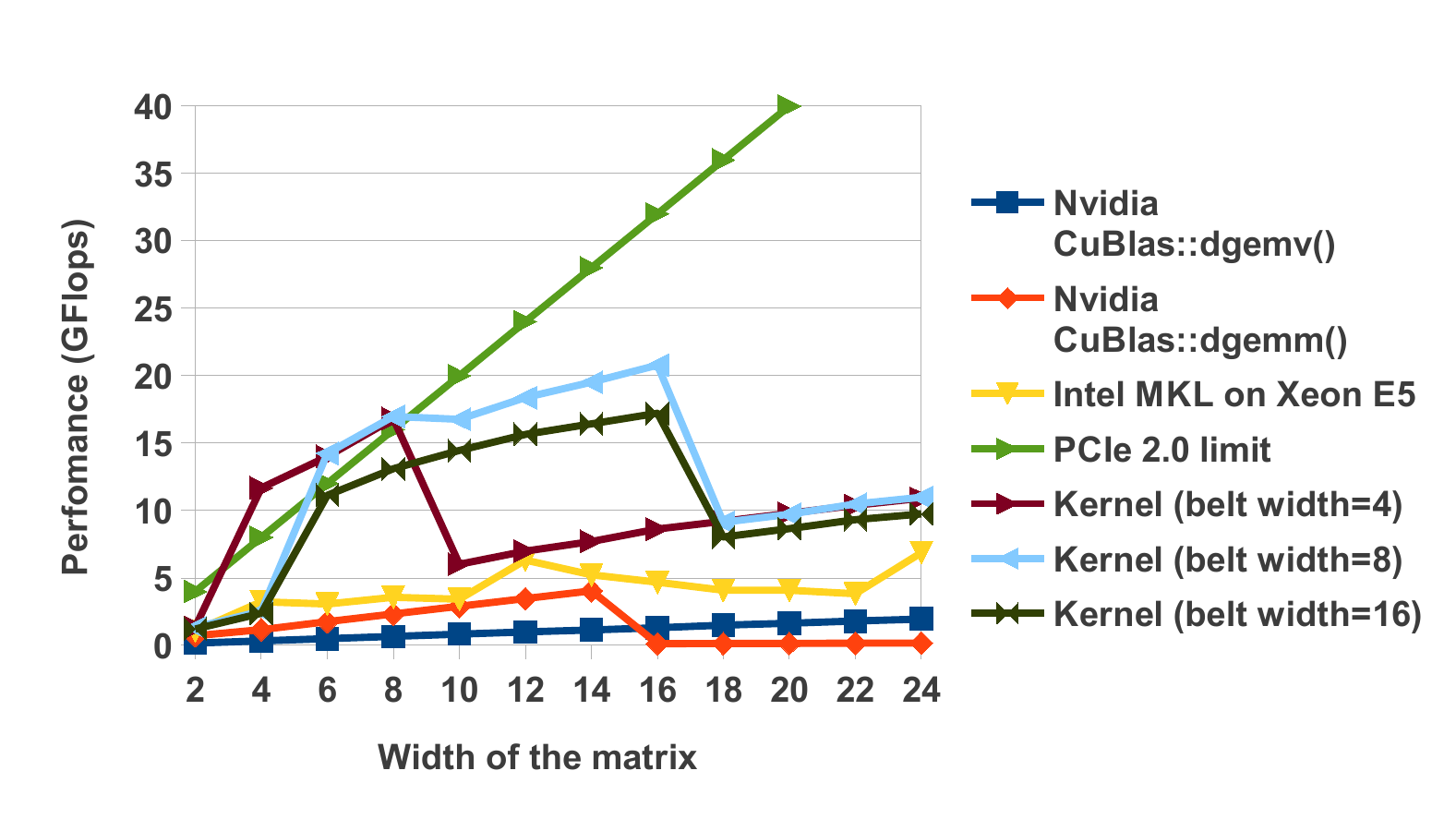}
  \end{center}
  \caption{K20, shuffle operation enabled: Performance of the presented \textit{gemv\_trans()} kernel with different belt widths is compared to the performance of the available implementations in case of matrix height $1e5$. 
Additionally, the PCIe limit is displayed as PCIe throughput limits the performance of the GPU acceleration if the transfer of the multiplier vector cannot be avoided.}
  \label{FigGemvk20shuffle1e5}
\end{figure}

\begin{figure}[t]
  \begin{center}
    \includegraphics[height=5cm]{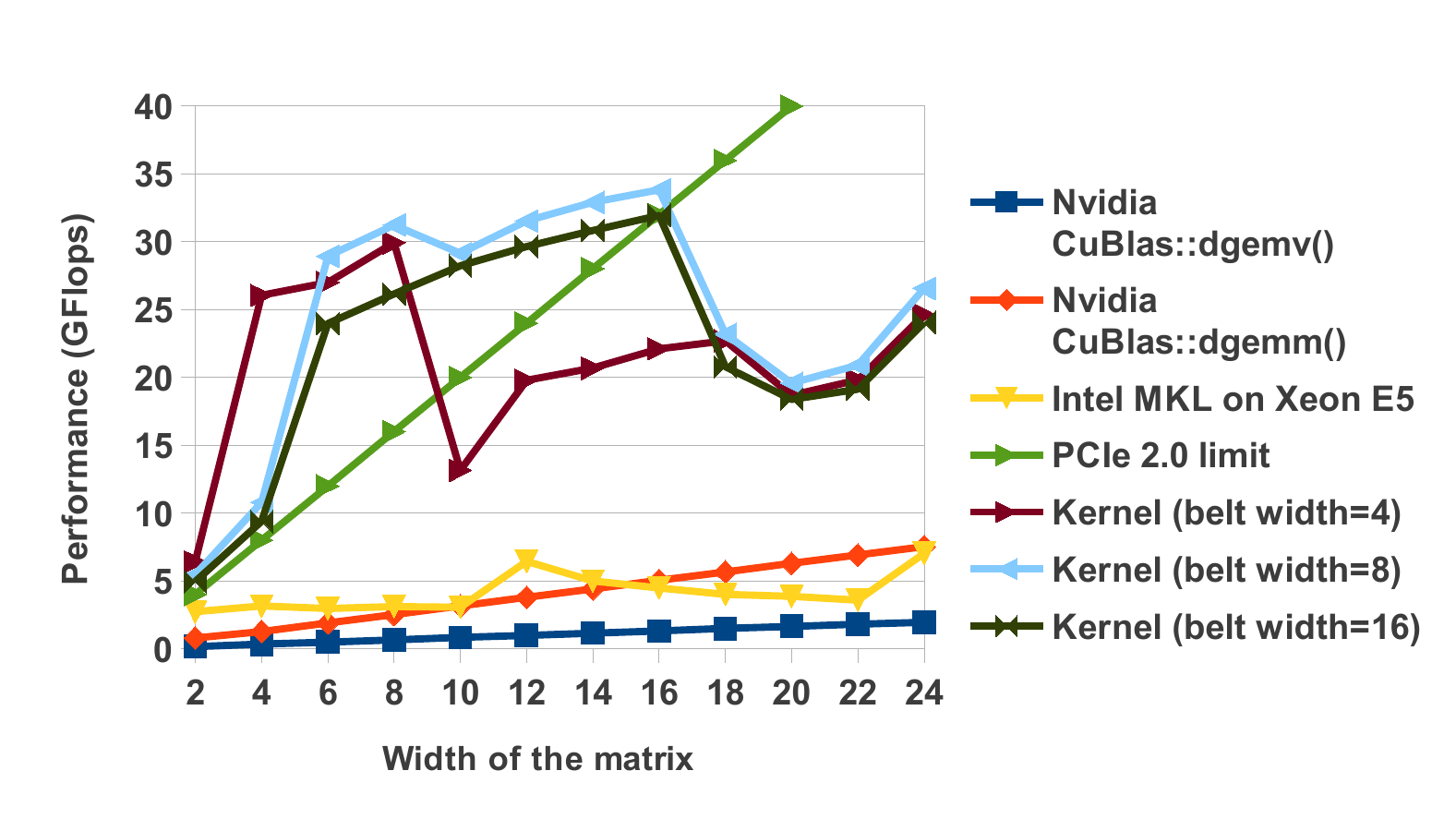}
  \end{center}
     \caption{Similar to Figure~\ref{FigGemvk20shuffle1e5} but for matrix height $5e5$.}
  \label{FigGemvk20shuffle5e5}
\end{figure}

\begin{figure}[t]
  \begin{center}
    \includegraphics[height=5cm]{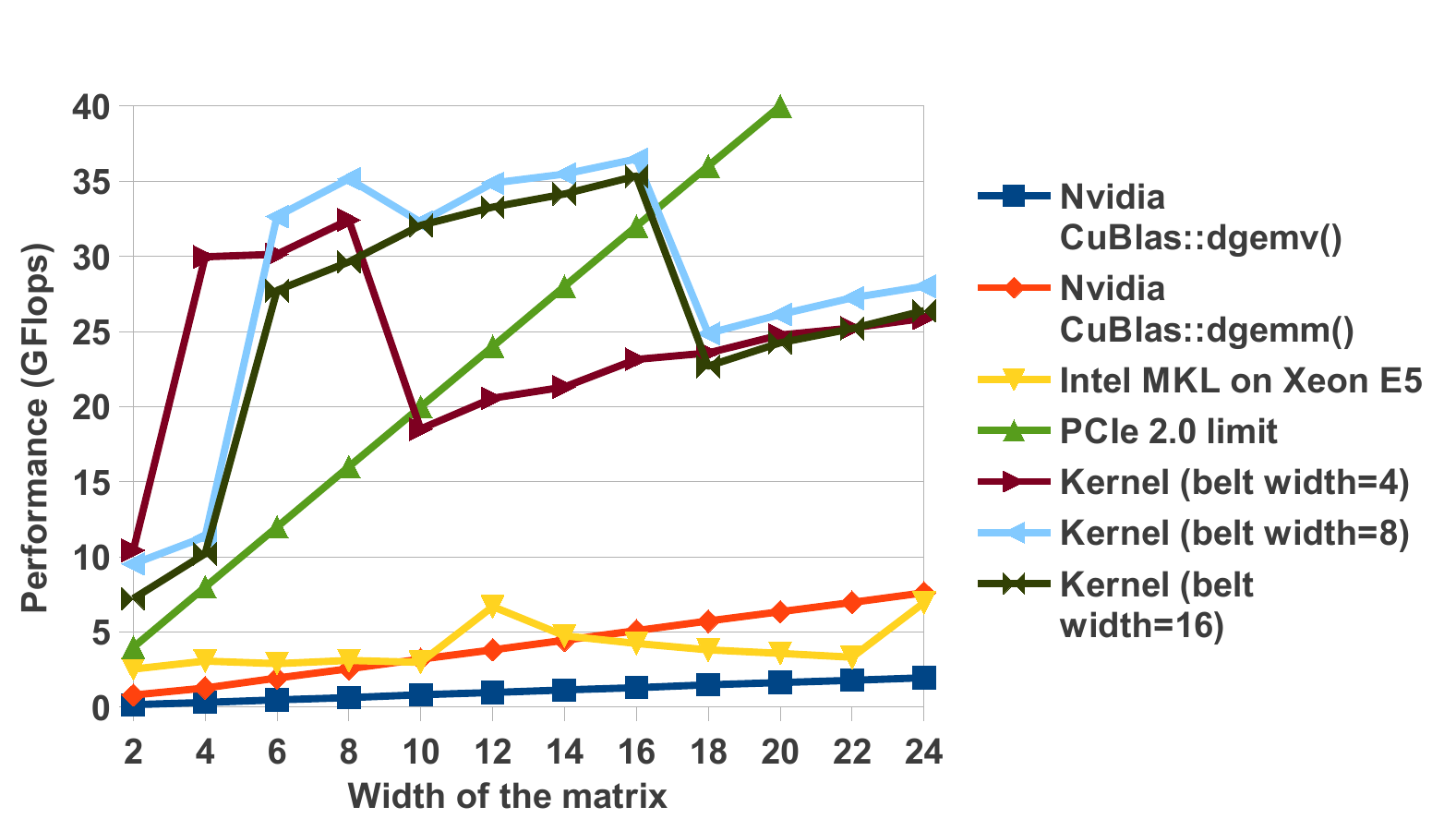}
  \end{center}
     \caption{Similar to Figures~\ref{FigGemvk20shuffle1e5} and~\ref{FigGemvk20shuffle5e5} but for matrix height $1e6$.}
  \label{FigGemvk20shuffle1e6}
\end{figure}

In case of gemv\_trans() both MKL and CuBlas libraries give poor performance for the special asymmetric matrix size (${\sim}5e5$x$20$) required in our application (see Figures~\ref{FigGemvGtx5e5},  \ref{FigGemvk20noshuffle}, \ref{FigGemvk20shuffle1e5}, \ref{FigGemvk20shuffle5e5} and \ref{FigGemvk20shuffle1e6}), therefore, a new CUDA kernel has been designed.
The presented results are measured without data communication, as in case of line 3 the multiplier vector is already in the GPU memory providing ideal acceleration. To estimate the performance of line 12, where the multiplier vector has to be loaded, the limiting factor of the PCIe 2.0 is also displayed. In this case the overall performance cannot exceed the PCIe limit even with overlapped communication. 
In both cases the estimation of the overall acceleration shall be carried out in an integral fashion, as in each iteration the thickness of the matrix is increased by one until a user defined limit ($20$ in the presented DMRG test-cases) is reached.

The basic idea of the new kernel (see Algorithm~\ref{algKernel}) can be summarized as follows. Each thread is associated with a column of the matrix. Each thread loads the corresponding vector element and multiplies the elements of the associated column. As threads of a warp load consecutive elements of the vector and the matrix, the coalesced reading is obvious. If the number of threads (grid size * thread block size) is less than the length of the matrix, each thread is associated with a new unprocessed column  (coalesced readings again) as long as there is any. After processing a new column each thread accumulates the results to the results of the first column. Finally, the accumulated results shall be summed across the threads, which can be efficiently done via a sum reduction~\cite{NVidiaProg} in shared memory.
If the belt is smaller than the width of the matrix, the whole procedure can be repeated (outer loop).

\begin{algorithm}[h!]
\caption{Proposed kernel for asymmetric gemv\_trans()}
\label{algKernel}
\begin{algorithmic}[1]
\Function{\_\_kernel\_\_ gemv\_trans(\\ mtx, mtx\_width, mtx\_len, vec)}{}
    \For {each belt}
        \For {(i = (blockIdx.x*blockDim.x) + threadIdx.x;  i $<$ mtx\_len; i += (blockDim.x*gridDim.x)) }
            \State priv = vec[i];
            \State \#define TMP(j) reg\#\#j  += mtx[i+ belt\_offset + j* mtx\_length] * priv;
            \State BOOST\_PP\_REPEAT(BELT\_SIZE,TMP);
	   \EndFor
	   \State ...
	   \State // Sum Reduction 
	   \If {(threadIdx.x == 0)} // Save results
	   \EndIf	
	\EndFor
\EndFunction
\end{algorithmic}
\end{algorithm}

The size of the shared memory requirement of a thread block, which is equal to the size of a thread block multiplied with the height of the belt, can be a limiting factor of the performance because in case of large shared memory usage less thread blocks can be assigned to one physical multiprocessor. In the presented measurements the optimal height of the belt has been investigated, however, even with the optimal height the performance decreases as the width of the matrix increases. 
For extreme, asymmetric matrices, which are used in our application, significant speed-up (x4-5) can be reached compared to the CuBlas library, however, as the matrix tends to be more symmetric the performance of the CuBlas::dgemm() operation (red line in the Figures) increases and exceeds the performance of the new kernel.

In case of the new Kepler architecture (K20), in which Streaming Multiprocessor has significantly more CUDA Cores than the SM of Fermi GPUs (GTX 570),  the per-multiprocessor warp occupancy shall be increased to use all the available cores~\cite{NVidiaKepler}. 
A new warp-level intrinsic called the shuffle operation can be used to decrease the shared memory requirement of the sum reduction algorithm to increase the occupancy.
In Figures~\ref{FigGemvk20shuffle1e5},~\ref{FigGemvk20shuffle5e5} and~\ref{FigGemvk20shuffle1e6} the results of the new kernel extended with the shuffle operation are displayed.
The height of the optimal belt is slightly increased as the shared memory request is decreased. Unfortunately, the shuffle operation provides only a small performance gain in case of our kernel (compare Figures~\ref{FigGemvk20noshuffle} and~\ref{FigGemvk20shuffle5e5}).


The performance of the kernel is mainly dominated by the speed of the coalesced reading of the matrix elements.
The gemv\_trans() operation is bandwidth limited in both CPU and GPU architectures, however, the memory bandwidth on GPU (e.g. GDDR5 in GTX 570: $152GB/s$ or GDDR5 in K20 $208GB/s$) is usually higher than on CPU (e.g. DDR3-1333 in dual channel with Core-i7: $21.2GB/s$ or DDR3-1066 in quad channel with Xeon E5: $34.1GB/s$).
The maximal memory throughput reached by the new kernel (measured with matrix size $16 \times 5e5$) was $114.7GB/s$, $134.8GB/s$ and $143.7GB/s$ on GTX 570, on K20 without shuffle and on K20 with suffle, respectively.


The results of the acceleration of the selected BLAS level 2 operations of the Davidson algorithm on the Xeon E5 + K20 architecture are summarized in Table~\ref{tabGemv}. (On the GTX 570 card the memory is too small to accelerate other operations besides the projection.) Line 3 is accelerated well as no extra communication is needed while the other operations are either limited by the PCIe or the DDR3 throughput.

\subsection{gemv()}

\begin{figure}[h!]
  \begin{center}
    \includegraphics[height=5cm]{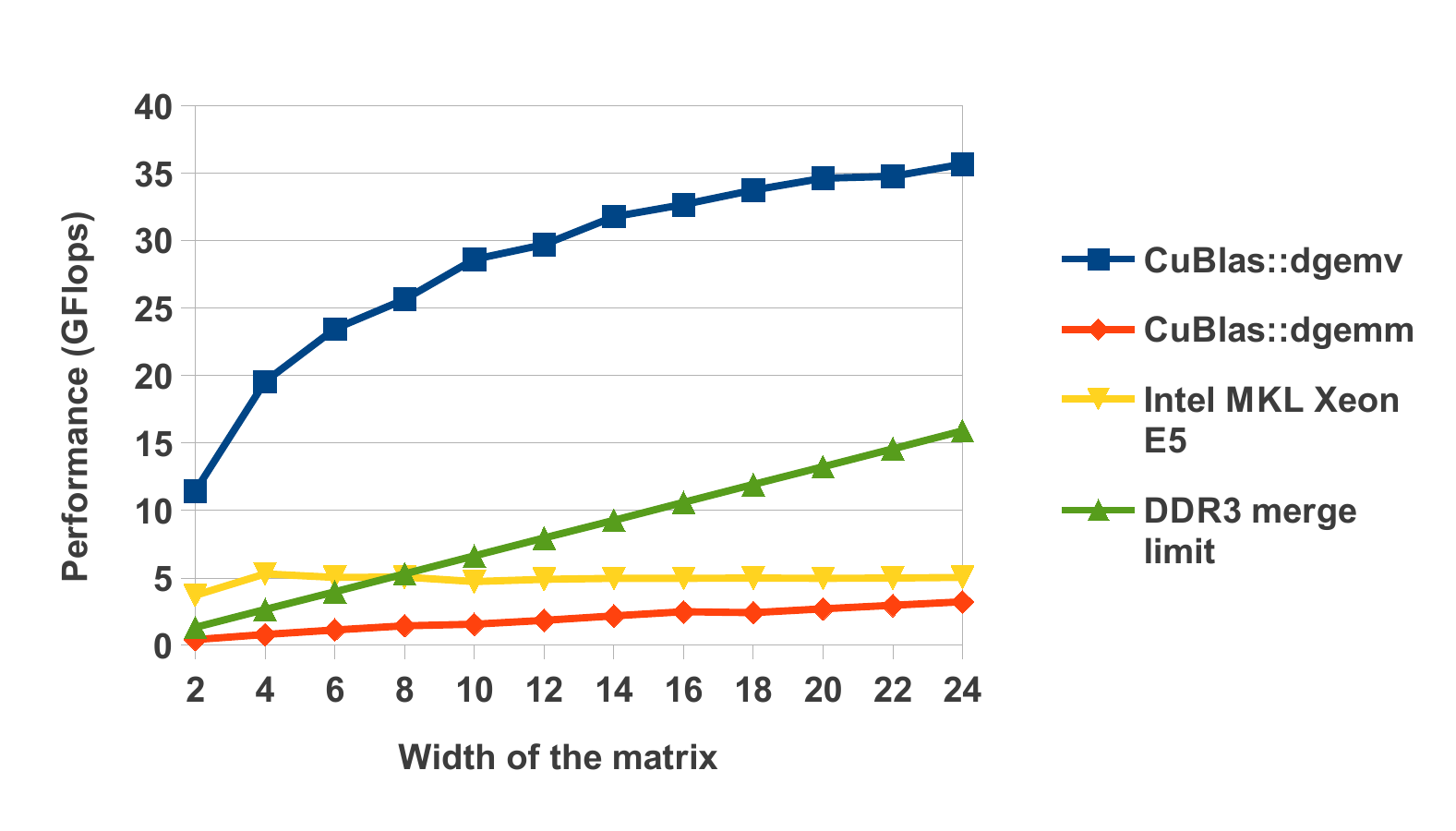}
  \end{center}
  \caption{Performance of the gemv\_normal() operation of the available implementations in case of K20. Additionally DDR3 limit is displayed, as in case of a CPU-GPU hybrid implementation, the merge operation is limited by the DDR3 throughput.}
  \label{FigGemvNormalK20}
\end{figure}

The gemv() operation can be efficiently accelerated by the standard CuBlas library even in case of asymmetric matrices (see Figure~\ref{FigGemvNormalK20}).
Unfortunately, merging of the CPU and GPU results is slow on one CPU thread and is the bottleneck of the acceleration. 
The implementation can be improved by multithreaded merging to enable quad channel memory or by computing everything on the GPU, however, this is not always possible.

\begin{table}[b]
\centering
\caption{Runtime of the accelerated matrix-vector operations of the Davidson algorithm (see Algorithm~\ref{algDav})}
\label{tabGemv}
\begin{tabular}{|l|l|c|c|c|}
\hline
\multicolumn{2}{|c|}{ }& Xeon E5	& Xeon E5 + K20	& speedup \\
\hline 
\multirow{3}{*}{\speci{Heisenberg\\ model}}
&Line 3& 20.21 &	4.64 &	4.36	\\ \cline{2-5}
&Line 5& 19.07 &	10.45&	1.83 \\ \cline{2-5}
&Line 12& 20.05 & 5.94	&3.38 \\ \cline{2-5}
&Line 13& 17.55 & 9.68	&1.81 \\ 
\hline 
\multirow{3}{*}{\speci{Hubbard\\ model}}
&Line 3& 113.80 & 22.66 & 5.02 \\ \cline{2-5}
&Line 5& 94.29 & 54.22 &	1.74 \\ \cline{2-5}
&Line 12 & 114.00 & 37.67 & 3.03 \\ \cline{2-5}
&Line 13& 87.29 & 50.21	&1.74 \\ 
\hline
\end{tabular} 
\end{table}

\section{Accelerating projection operation}\label{SecAccelProjection}

The acceleration of the independent $(AX)B^T$ operations implementing the projection operation is based on the observation that $A$ and $B$ matrices are already available before the Davidson algorithm starts and do not change during the Davidson iterations. 
The necessary $(AX)B^T$ operations are described by a list of \textit{operation records} in which each record contains all the necessary information to compute an operation like Equation~\ref{EgProjectionDecomposed}. 
For example, it stores information from which segment of $X$ (\textit{input}) to which segment of $X'$ (\textit{output}) the operation transforms.

The host side algorithm to handle the operation records is summarized in Algorithm~\ref{alg2}.
During the construction of the operation records the workload associated to each output is computed.
(Multiple operations can use the same input or write the same output segment.)
Next, the operation records are partitioned between CPU and GPU based on the performance ratio of the two architectures. 
To avoid merging of outputs all operation records corresponding to the same output shall be computed on the same architecture, however, to create a balanced workload partitioning, this is not always possible.
During partitioning the output associated to the largest workload is selected for GPU iteratively as long as the desired workload ratio is not exceeded.
If the reached workload ratio is far from the desired, the operation records of the output associated to the next largest workload are partitioned between the two architectures.

\begin{algorithm}[h!]
\caption{Host side algorithm to handle the operation records}
\label{alg2}
\begin{algorithmic}[1]
\State Create operation records and determine the workload (FLOP) for each output.
\State Partition the operation records between CPU and GPU based on their performance ratio and the output workload statistics.
\State Selects scheduling strategy for the operations to be computed on GPU.
\State Set-up the workload for GPU based on the selected strategy.
\end{algorithmic}
\end{algorithm}

\begin{figure}[b]
  \begin{center}
    \includegraphics[height=5cm]{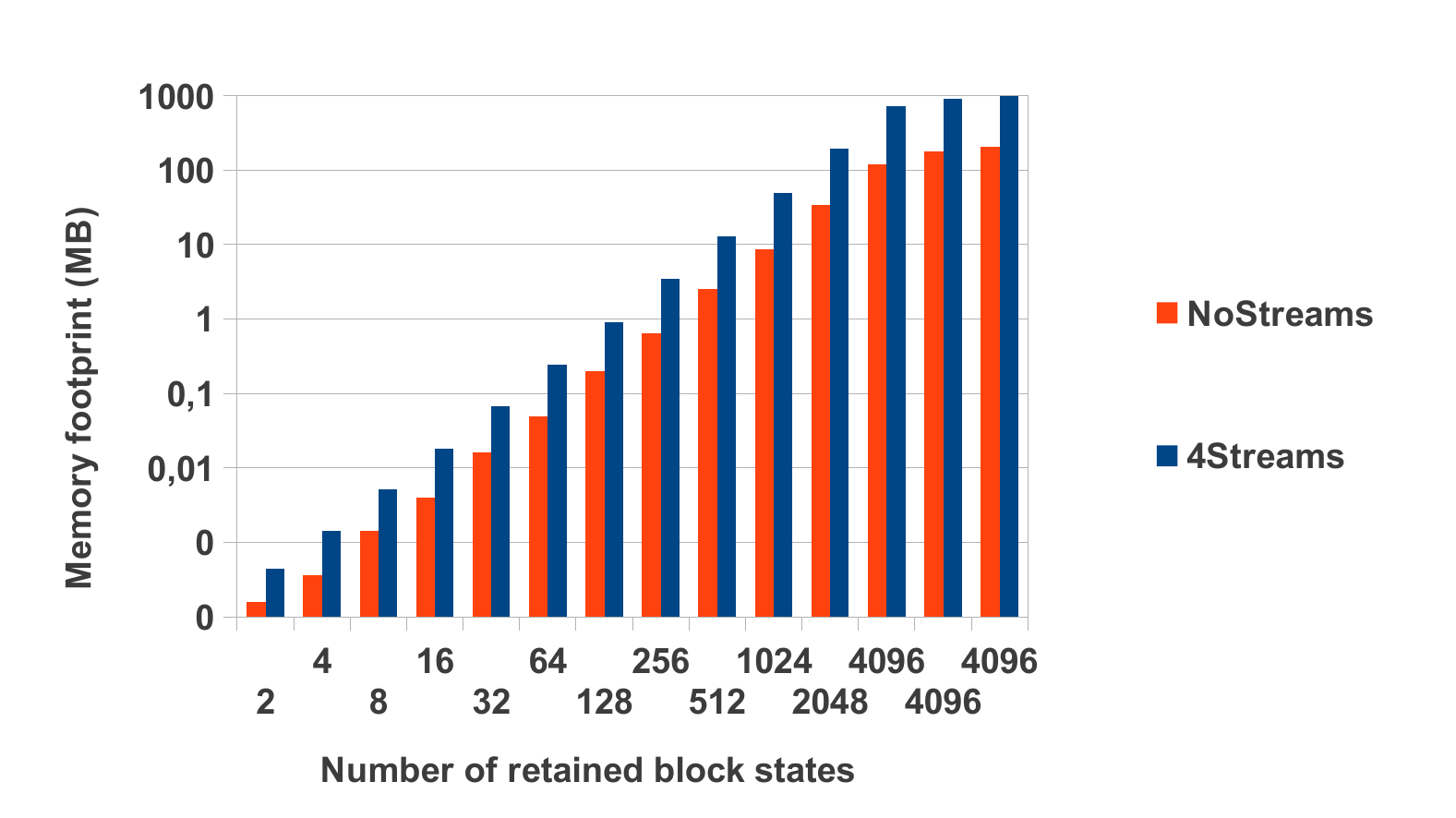}
  \end{center}
  \caption{GPU memory footprints of the two strategies are compared in case of the Heisenberg model.}
	\label{FigDifferentStrategies}
\end{figure}

After partitioning the proper scheduling strategy is selected based on the memory requirements of the operation records.
Three different strategies can be selected for three different uses cases, however, currently only the first two, more complex strategies have been implemented and tested.
The first strategy (\textit{4Streams}) is designed for small problem size, when all $A$, $B$, $X$, $X'$ matrices and temporary matrices $T$ for storing intermediate results can be held in the GPU memory.
The second strategy (\textit{NoStreams}) is designed for medium-sized problems, where all $A$, $B$ and $X'$ matrices can be stored in GPU memory, but from $X$ and $T$ only the processed matrices are allocated.
In case of extra-sized problems a third strategy (\textit{NoStreamAndStorage}) can be designed in which even $A$ and $B$ matrices cannot be fully stored in the GPU memory.
The memory footprint of the matrices in case of different strategies are shown in Figure~\ref{FigDifferentStrategies}. 
In the demonstrated examples the second strategy is sufficient for all DMRG iterations as both GPU cards have enough memory.

The $(AX)B^T$ operations are implemented using the cuBLAS libary~\cite{NVidiaCuBLAS}, which is a BLAS implementation dedicated for Nvidia GPUs. In the demonstrated strategies two important features of the GPUs are exploited, which are provided via the CUDA driver~\cite{NVidiaCuda} and also accessible through the cuBLAS library. 
The first feature is that multiple CUDA kernels can be executed simultaneously on the GPU, while the second feature is that memory I/O operations can be executed in the background. From the aspect of programming both features can be accessed via the CUDA streams. Streams are sequences of operations that execute in issue-order, but operations in different streams may run concurrently or interleaved.

In the 4Streams strategy there is enough GPU memory to execute several $(AX)B^T$ simultaneously. 
One stream is created for each output and operations corresponding to a given output are assigned to the same stream to avoid interference.
For each stream a sufficiently large temporary matrix is allocated to store the temporary result of $AX$.

CUDA operations are dispatched to \textit{hardware queues} in issue order~\cite{NVidiaCuda}.
To enable asynchronous concurrent kernel execution in CUDA environment memory transfers and kernels shall be issued in a depth-first order.
Inside the engine (kernel) queue an operation is dispatched if all preceding calls in the same stream have been completed and all preceding calls of the same queue have been dispatched.
Consequently, to avoid blocking calls kernels of the same streams shall not be issued immediately after each other.
As one $(AX)B^T$ operation consists of two kernels, the kernel calls shall be separated and interleaved with kernels of operations of other streams.
To reach four parallel streams (hence the name of the strategy)  kernels from four different stream shall be interleaved.
\begin{algorithm}[h!]
\caption{Grouping operation records in case of 4Streams strategy}
\label{alg4StreamsOrder}
\begin{algorithmic}[1]
\Function{orderAndGroupRecords}{$records$, $maxstream$}
\State Sort $records$ by input frequency.
\State setVisitedRecords.clear()
\For {each record $i$}
	\If {$i.stream$ $\in$ setVisitedRecords}
		\For {each record $j$ following $i$}	
			\If {$j.stream$ $\not\in$ in setVisitedRecords}
				\State swap($i$,$j$) and break
			\EndIf
		\EndFor
	\EndIf
	\If {$i.stream$ is $\not\in$ in setVisitedRecords}
		\State vecGroup.last().insert($i$)
		\State setVisitedRecords.insert($i.stream$)
		\If {setVisitedRecords.size()=$maxstream$}
			\State vecGroup.add(new Group)
			\State setVisitedRecords.clear() 
		\EndIf	
	\Else
		\State vecGroup.add(new Group)
		\State vecGroup.last().insert($i$)
		\State setVisitedRecords.clear()
		\State setVisitedRecords.insert($i$)
	\EndIf
\EndFor
\Return $vecGroups$
\EndFunction
\end{algorithmic}
\end{algorithm}

Overlapping of the transfer time of input segments with kernel execution makes further constraints on the order of the operation records: only those operation records shall be issued which use already loaded input segments.
To be able to interleave different streams, it is favorable to load the input segment first which is used by the most streams.
\begin{algorithm}[h!]
\caption{Dispatching operation records}
\label{alg4StreamsIssue}
\begin{algorithmic}[1]
\For {each group $g$}
	\For {each record $i$ in $g$}
		\State Init copy of input segment $X_i$ (when first used)
		\State Init $T_i=(A_iX_i)$
	\EndFor
	\For {each record $i$ in $g$}
		\State Init $X_i'=T_iB_i^T$
	\EndFor
\EndFor
\For {each output segments of $X'$} Init copy back.
\EndFor
\end{algorithmic}
\end{algorithm}

In case of 4Streams strategy the reordered operation records are grouped (see Algorithm~\ref{alg4StreamsOrder}) such a way that kernels belonging to the same group can be interleaved (see Algorithm~\ref{alg4StreamsIssue}) in execution.
First, operation records are sorted to load the more frequently used input segments earlier.
Then, the records are iterated and each record is potentially swapped backwards to create groups of four consecutive operations belonging to four different streams. In practice some technical constraints have been added to slightly alter swapping behavior, which is not discussed here for the sake of simplicity.

CUDA operations are launched according to the operation records as summarized in Algorithm~\ref{alg4StreamsIssue}. For the sake of brevity the synchronization between the streams is not shown as it can be implemented with CUDA events in a straightforward way.
The same code can be used for both strategies as the NoStreams strategy can be represented by groups which contain only one operation record.
In case of NoStreams strategy the preparation of the operation records is much simpler and contains only the sorting by input frequency to reach I/O overlap with computation.

\begin{figure}[h!]
  \begin{center}
    \includegraphics[height=5cm]{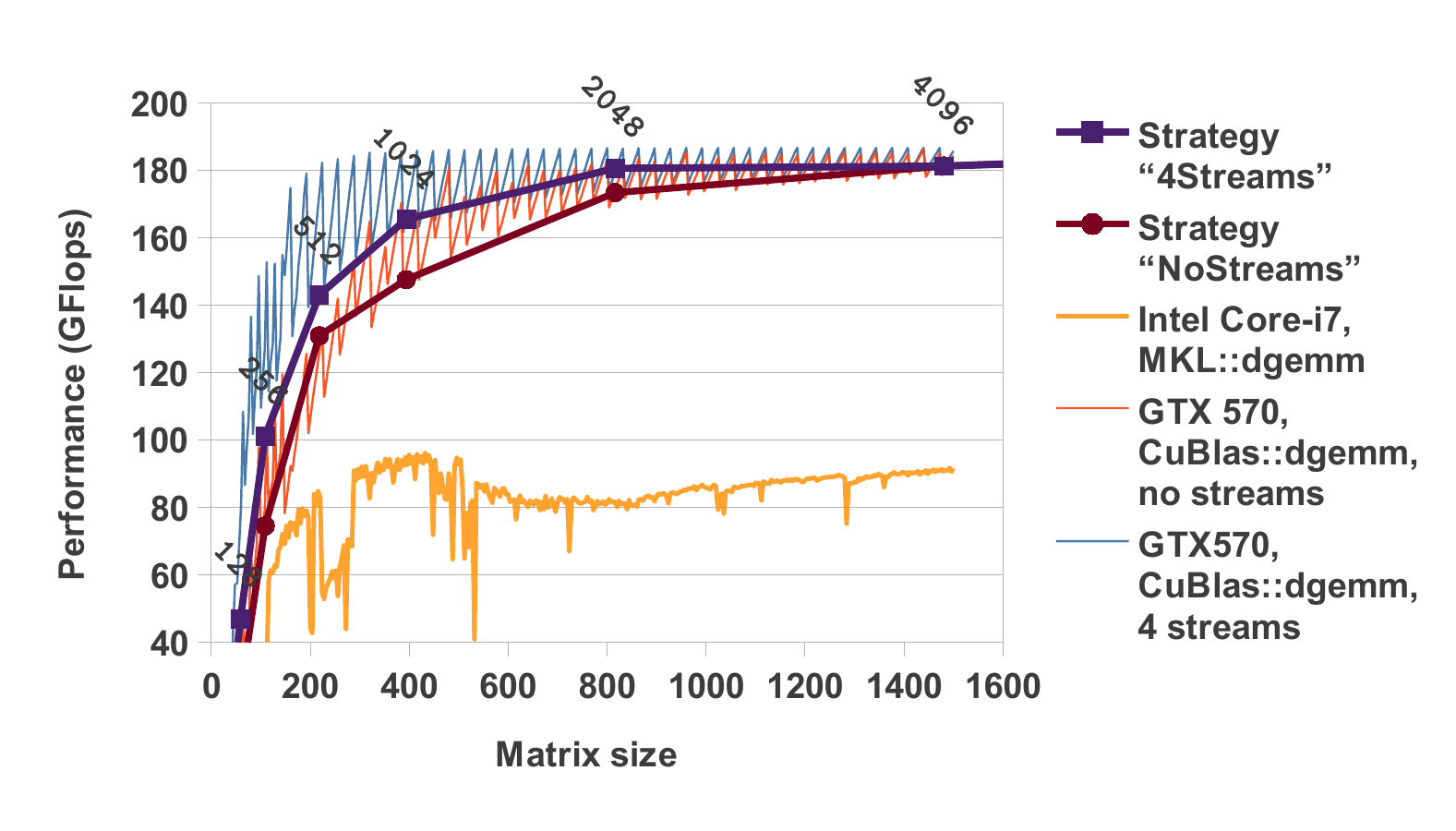}
  \end{center}
  \caption{GTX 570, Heisenberg model: Performance of the two strategies is compared. Additionally, the performance of CuBLAS and MKL dgemm() in reference measurements is displayed as the function of matrix size. Labels indicate the number of retained block states at the displayed DMRG iterations.}
  \label{FigStratGtxHeis}
\end{figure}
\begin{figure}[h!]
  \begin{center}
    \includegraphics[height=5cm]{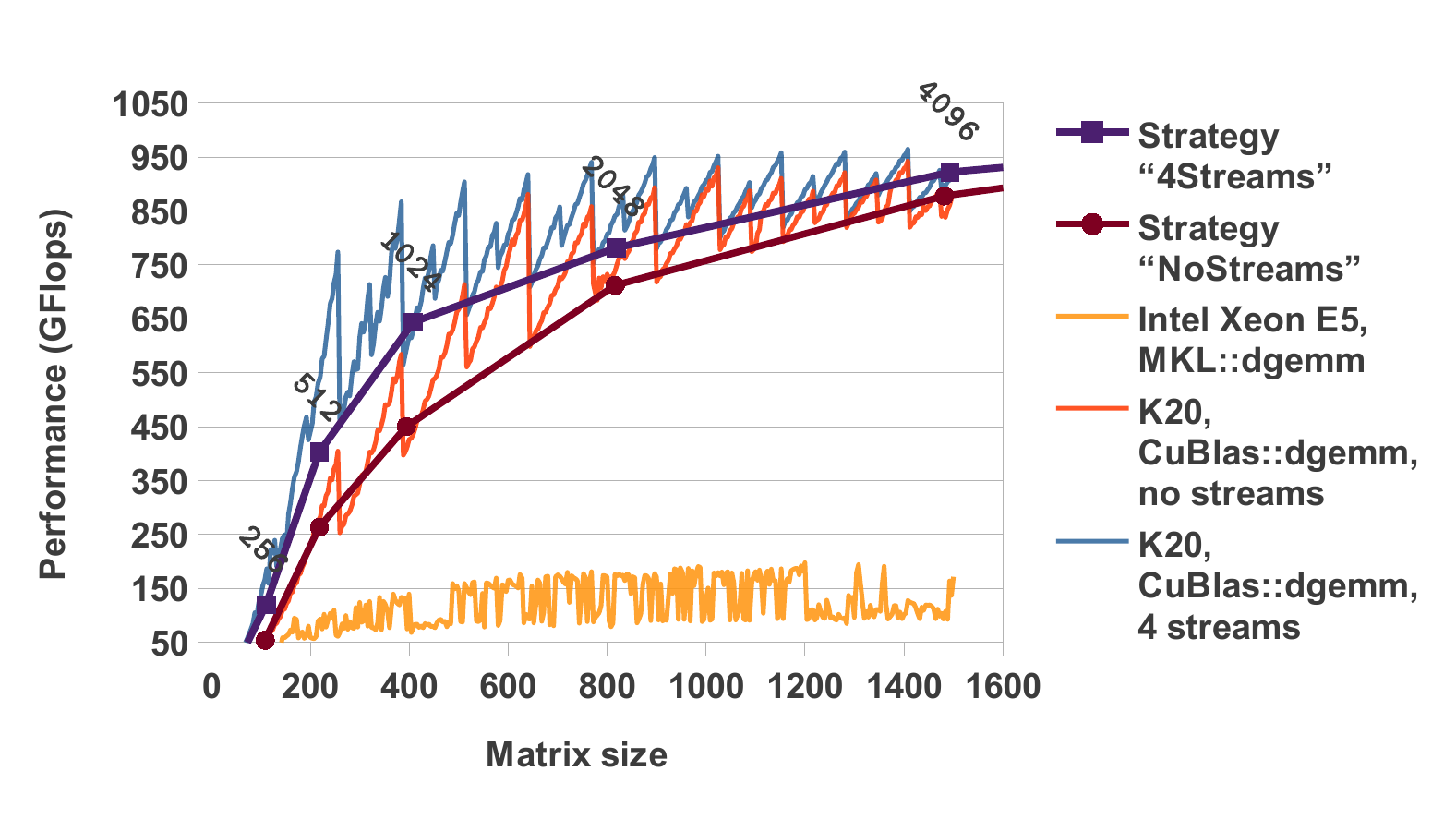}
  \end{center}
   \caption{Similar to Figure~\ref{FigStratGtxHeis} but on K20 architecture.}
  \label{FigStratK20Heis}
\end{figure}

The performance of the two strategies is compared in Figures~\ref{FigStratGtxHeis} and~\ref{FigStratK20Heis}.
Significant improvement can only measured at medium sized matrices ($100$-$800$ for GTX 570 and $100$-$1500$ for K20), in which case several operations shall be executed concurrently to keep all the CUDA cores busy required for high performance.
Slightly bigger gain can be observed in case of K20 GPU which has $2496$ Kepler CUDA cores as opposed to GTX 570 having only $480$ Fermi CUDA cores.
Operations on large matrices (${\sim}1500$ for GTX 570 and ${\sim}3000$ for K20) provide enough work for each CUDA core to approach the theoretical maximum double performance ($180$ GFlops for GTX570 and $1.17$ TFlops for K20) without streams.

The two strategies are also compared by the run-time of the simulated models in Table~\ref{tabStrat}.
In case of K20 the concurrent kernel execution has a slightly greater benefit, however, in both models operations on larger matrices, where concurrency has no benefit, dominates the run-time. 
In models where more symmetries are enabled the size of the matrices tends to be smaller, consequently, in these models the concurrency also tends to be more effective.

\begin{table}
\centering
\caption{Total time of strategies is compared}
\label{tabStrat}
\begin{tabular}{|c|c|c|c|c|}
\hline 
 &Model& \speci{NoStream\\(sec)} & \speci{4Streams\\(sec)} & decrease \\ 
\hline 
\multirow{2}{*}{GTX570}&Heisenberg & 671.54 & 652.58 & 2.82\% \\ 
\cline{2-5}
&Hubbard & 2980.27 & 2957.82 & 0.75\% \\ 
\hline 
\multirow{2}{*}{K20}&Heisenberg & 244.67 & 227.33 & 7.09\% \\ 
\cline{2-5}
&Hubbard & 1056.33 & 1012.56 & 4.14\% \\ 
\hline 
\end{tabular} 
\end{table}

\begin{figure}[t]
  \begin{center}
    \includegraphics[height=5cm]{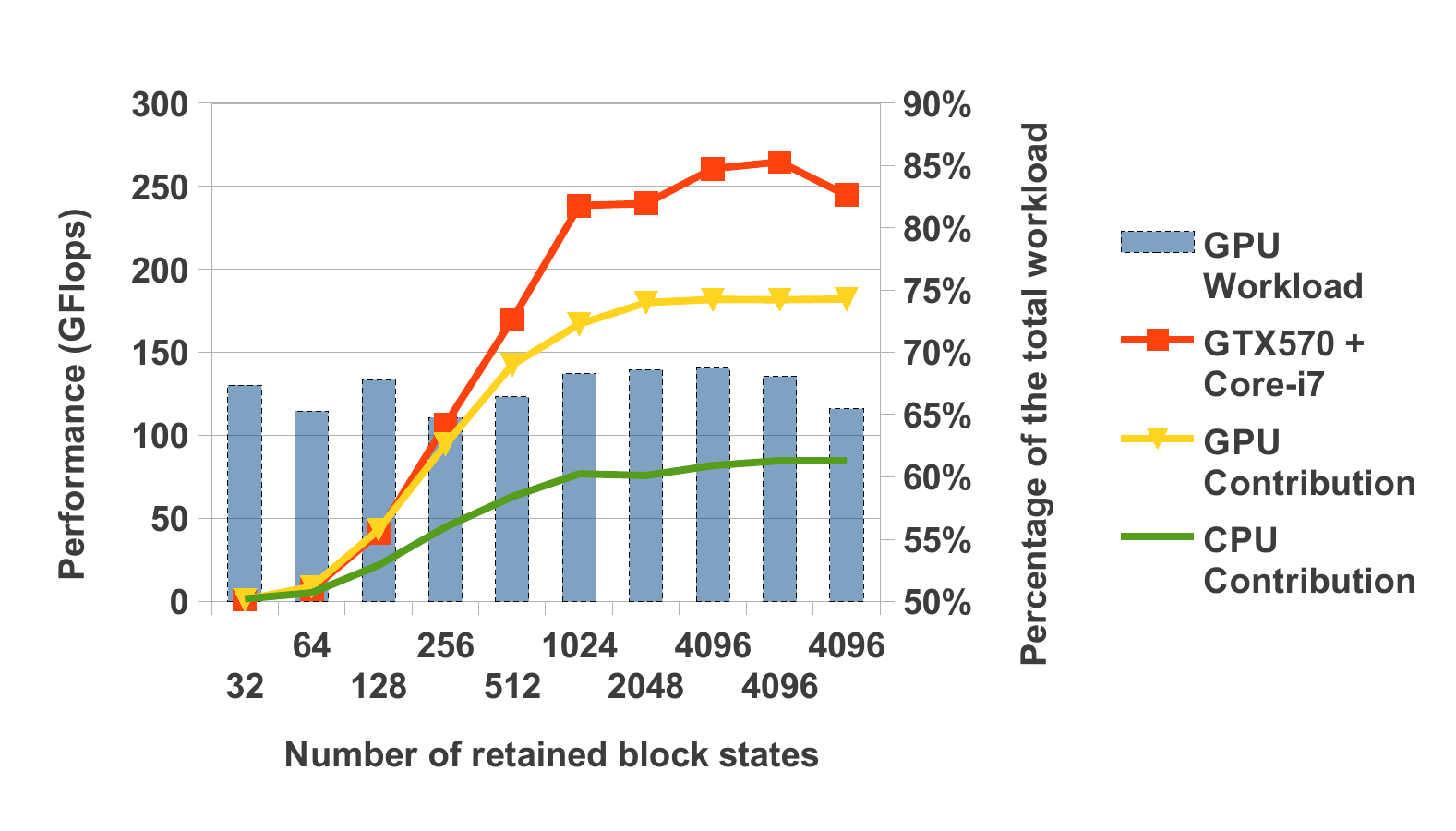}
  \end{center}
  \caption{GTX 570, Heisenberg model: Performance results of the hybrid CPU-GPU acceleration of the projection operation. Blue bars associated to the secondary vertical axis indicate the ratio of the current GPU workload.}
  \label{FigHybGtxHeis}
\end{figure}

\begin{figure}[t]
  \begin{center}
    \includegraphics[height=5cm]{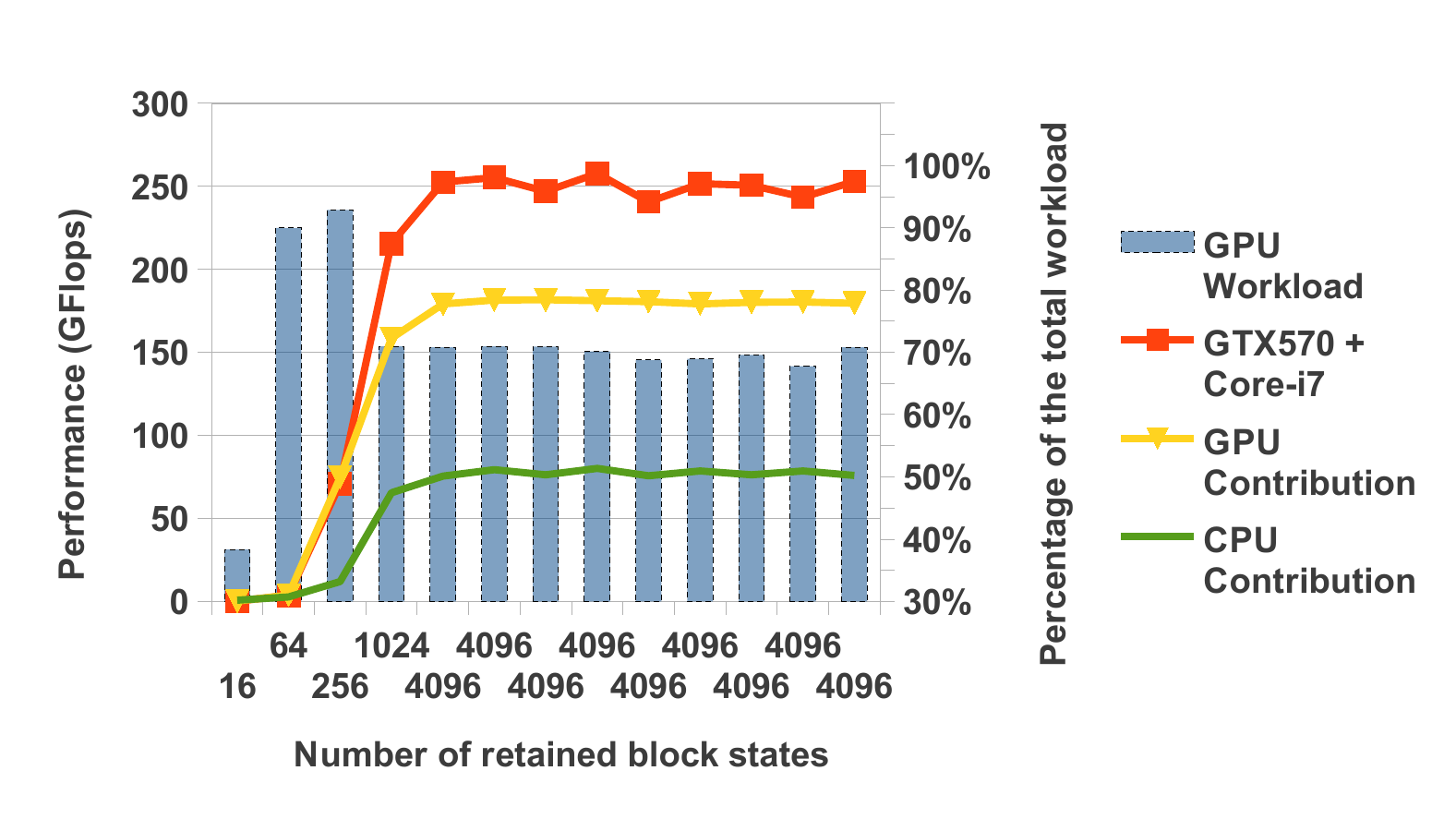}
  \end{center}
 \caption{Similar to Figure~\ref{FigHybGtxHeis} but for the Hubbard model on GTX 570.}
  \label{FigHybGtxHub}
\end{figure}

\begin{figure}[t]
  \begin{center}
    \includegraphics[height=5cm]{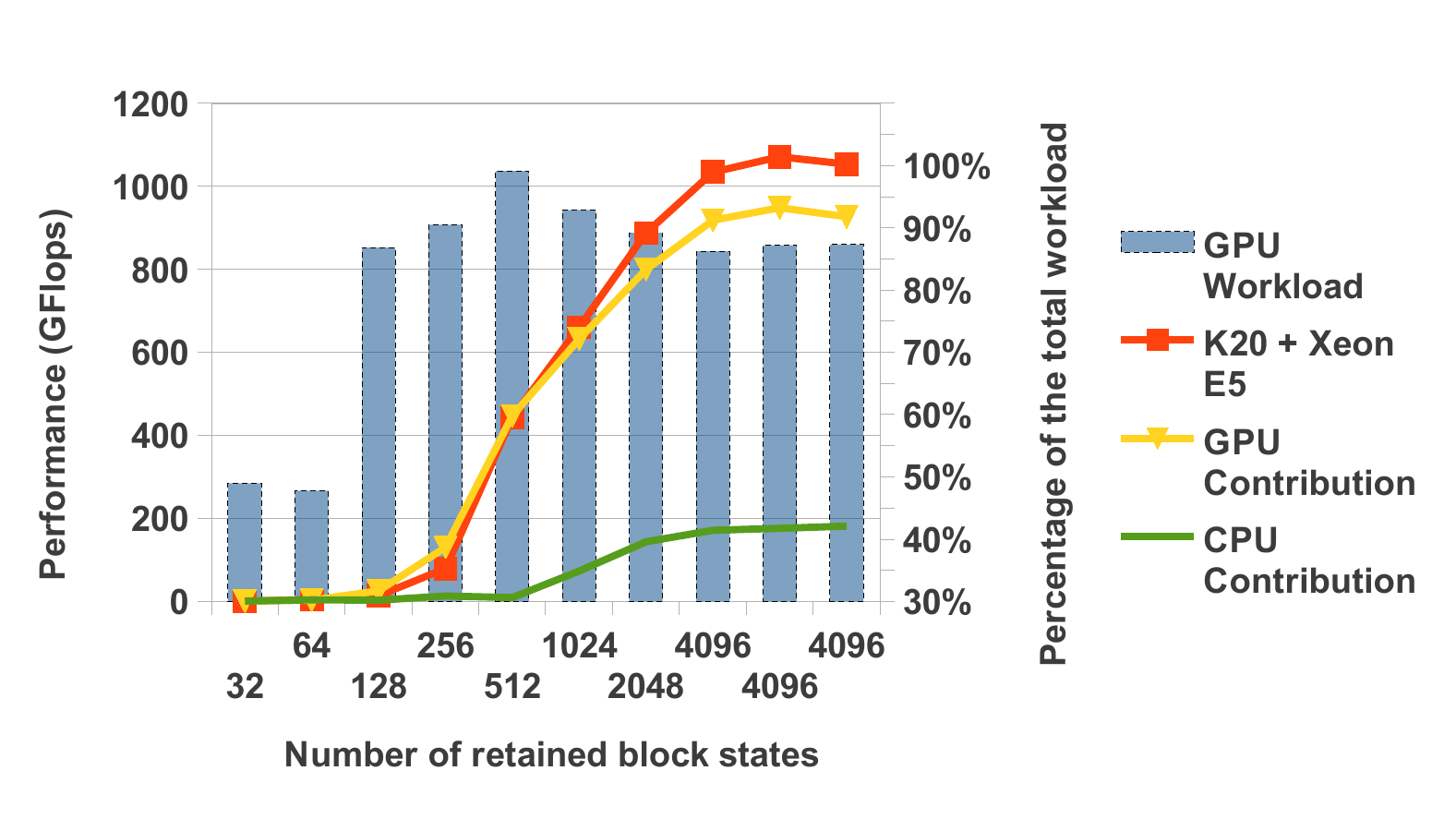}
  \end{center}
   \caption{Similar to Figures~\ref{FigHybGtxHeis} and~\ref{FigHybGtxHub} but for the Heisenberg model on K20.}
  \label{FigHybK20Heis}
\end{figure}

\begin{figure}[t]
  \begin{center}
    \includegraphics[height=5cm]{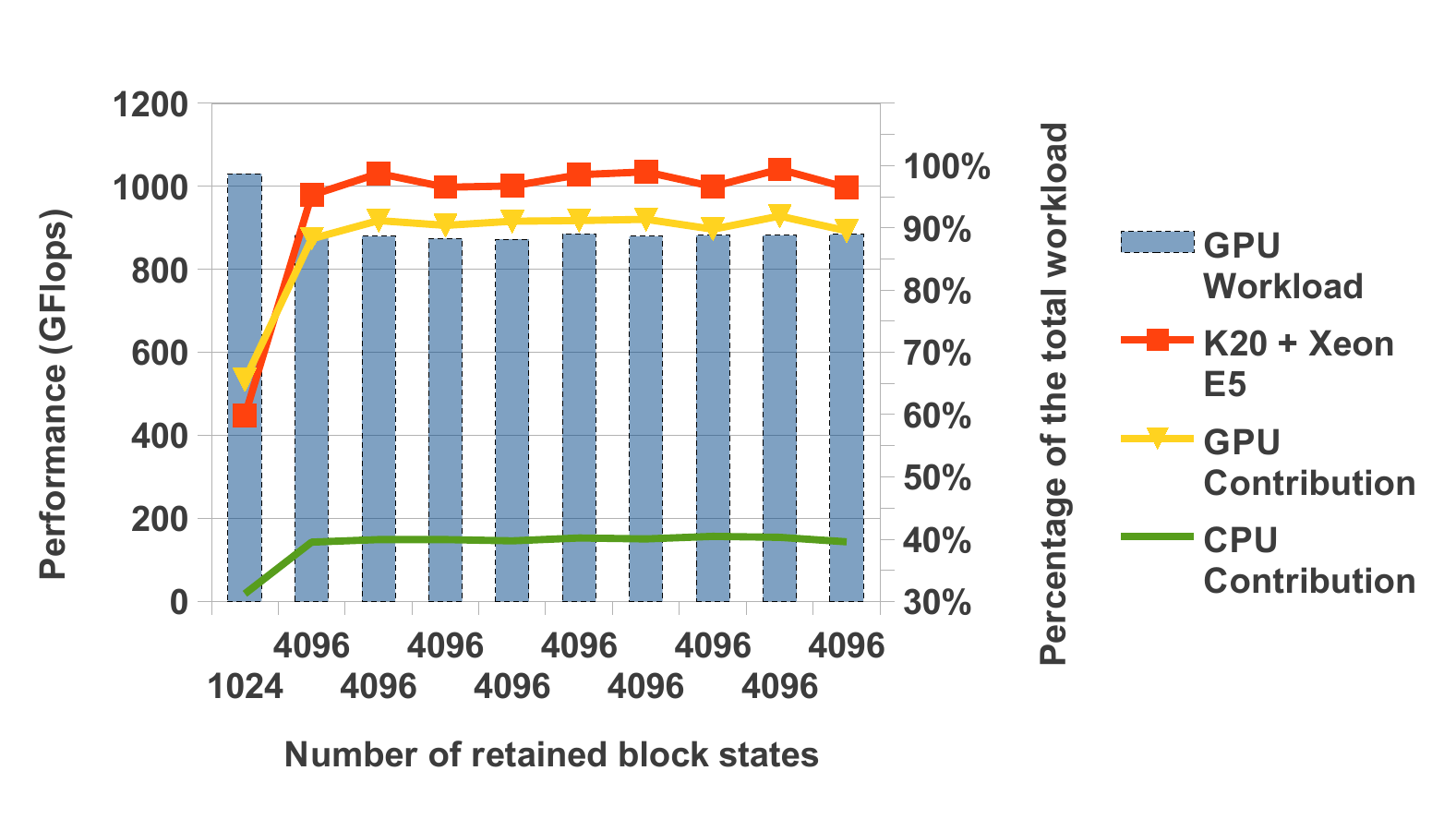}
  \end{center}
   \caption{Similar to Figures~\ref{FigHybGtxHeis},~\ref{FigHybGtxHub} and~\ref{FigHybK20Heis} but for the Hubbard model on K20.}
  \label{FigHybK20Hub}
\end{figure}

The performance results of the full projection computation including both CPU and GPU computations are shown in Figures~\ref{FigHybGtxHeis},~\ref{FigHybGtxHub},~\ref{FigHybK20Heis} and~\ref{FigHybK20Hub}.
The quality of the acceleration is highly affected by the applied workload ratio which depends on the performance ratio of CPU and GPU at the given matrix size.
In the configuration file different ratios can be set for different matrix sizes and in each DMRG iteration the user-defined ratio is selected according to the average matrix size of the operation records.
If the workload is properly distributed $257.8$ GFlops  (${\times}3.2$ speed-up) and $1071.1$ GFlops (${\times}6.1$ speed-up) can be reached on GTX 570 and on K20, respectively.


\section{Limits of FPGA implementation}\label{SecFPGA}

Field programmable gate arrays (FPGAs) are programmable integrated circuits used to realize reconfigurable digital circuits via configurable logic resources and routing. Although originally developed for telecommunication and digital signal processing applications novel high performance FPGAs (featuring high-speed embedded resources, plentiful on-chip memory, high-level programming tools and significantly lower power consumption than CPU or GPU) are promising candidates for high performance computing (HPC).

To estimate the performance of an FPGA implementation of the DMRG method the acceleration of the projection operation expressed as a series of dense matrix multiplications (see Equation~\ref{EgProjectionDecomposed}) shall be investigated.
The floating-point matrix-matrix multiplication was already implemented~\cite{Kumar2010} on FPGA very efficiently using the rank-1 update scheme. Kumar et al. demonstrated that the performance is not limited by the PCIe bandwidth, which connects the FPGA to the host CPU, and nearly full utilization of the processing elements can be reached.

The idea behind the rank-1 update approach is that instead of inner products between the rows of the left matrix and the columns of the right matrix outer products between the columns of the left matrix and the rows of the right matrix are carried out and resulting matrices are summarized. The advantage of the approach is that instead of multiply–accumulate operations (MACCs) multiply-add operations (MADDs) are used, which are independent from each other and can be pipelined to reach high processing element utilization.

Assuming the previously described design the processing elements can be fully utilized and the following best-case estimations can be made according to the area requirements of the floating-point units.
In an ideal case at most 114 or 193 multiply-add units can be implemented on the largest
Virtex-6 ( XC6VSX475T ) and Virtex-7 (XC7VX1140T) FPGAs, respectively.
The estimated clock frequency of the architectures are 437.82 and 443.65MHz which would result in 99.82 and 171.2 GFLOPS computing performance.
This performance achievement can be compared to the performance of the mid-range GTX 570 GPU used in the paper. 
As the development time in case of FPGA is still much longer than in case of GPU and the high-end GPUs could significantly outperform the FPGA in this problem class, the GPU architecture is the better candidate for the acceleration.

\section{Implementation Results}\label{SecImp}

The implemented algorithm has been tested both on a mid-range (Intel Core-i7 2600 3.4 GHz CPU + NVidia GTX 570 GPU)  and on a high-end configuration (Intel Xeon E5-2640 2.5 GHz  CPU + NVidia K20 GPU); the results are displayed in Table~\ref{tabFinalGtx} and~\ref{tabFinalK20}, respectively.
All CPU-only measurements have been executed with multithreading enabled (4 threads on Core-i7 and 6 threads on Xeon E5).
The mid-range configuration with GPU is approximately $2.3$-$2.4$ times faster than without GPU, while the high-end configuration is accelerated by $3.4$-$3.5$ times using the GPU.
However, a change from a mid-range, multithreaded CPU to a high-end CPU+GPU configuration can produce $6.5$-$7$ times acceleration.

\begin{table}[h!]
\centering
\caption{Heisenberg model: final timings compared}
\label{tabFinalGtx}
\begin{tabular}{|l|c|c|c|}
\hline 
 & \multirow{2}{*}{Time(sec)} & \multicolumn{2}{|c|}{Speed-up compared to}  \\ 
 \cline{3-4}
 &&Core-i7&Xeon E5 \\
\hline 
Core-i7 & 1489.64 & 1 & 0.53 \\ 
\hline
Core-i7 + GTX 570 & 652.58 & 2.28 & 	1.21 \\ 
\hline 
Xeon E5 & 789.65 & 1.89 & 1 \\ 
\hline
Xeon E5 + K20 & 227.33 & 6.55 & 	3.47 \\ 
\hline 
\end{tabular} 
\end{table}

\begin{table}[h!]
\centering
\caption{Hubbard model: final timings compared}
\label{tabFinalK20}
\begin{tabular}{|l|c|c|c|}
\hline 
 & \multirow{2}{*}{Time(sec)} & \multicolumn{2}{|c|}{Speed-up compared to}  \\ 
 \cline{3-4}
 &&Core-i7&Xeon E5 \\
\hline 
Core-i7 & 7210.72 & 1 & 0.48 \\ 
\hline
Core-i7 + GTX 570 & 2957.82 & 2.44 & 1.16 \\ 
\hline 
Xeon E5 & 3433.16 & 2.10 & 1 \\ 
\hline
Xeon E5 + K20 & 1012.56 & 7.12 & 3.39 \\ 
\hline 
\end{tabular} 
\end{table}
\begin{table}[h!]
\centering
\caption{Model comparison in case of Xeon E5 + K20.}
\label{tabModelComp}
\begin{tabular}{|l|c|c|c|}
\hline 
 & Heisenberg & Hubbard & ratio \\
\hline 
Time(s) & 244.67 & 1067.89 & 4.36 \\ 
\hline
Flop & 1.22E+014 & 4.89E+014 & 4.01 \\ 
\hline 
Max $H_{SB}$ size & 12.24E+06 & 15.32E+06 & 1.25 \\ 
\hline
Max Sector size & 4.00E+06 & 3.47E+06 & 0.87 \\ 
\hline 
\speci{Average number of\\ sectors} & 9.36 & 50.71 & 5.42 \\ 
\hline 
Max matrix size  & 1704.23 & 1145.24 & 0.67 \\ 
\hline 
\speci{Peak GPU memory\\ footprint} & 950.47 & 1155.48 & 1.22 \\ 
\hline 
\speci{Average number of\\ Davidson iterations using\\ random starting vector}& 60.79 & 122.43 & 2.01 \\ 
\hline 
\end{tabular} 
\end{table}

To support the comparison of the results of the two investigated models the key parameters affecting computational complexity are summarized in Table~\ref{tabModelComp}. Using the same number of retained block states the Hubbard model has larger values for all key parameters except the maximum sector size and the maximum matrix size. In case of the Hubbard model more symmetries are exploited which results in smaller sectors and, consequently, smaller matrices.

\begin{figure}[b]
  \begin{center}
    \includegraphics[height=5cm]{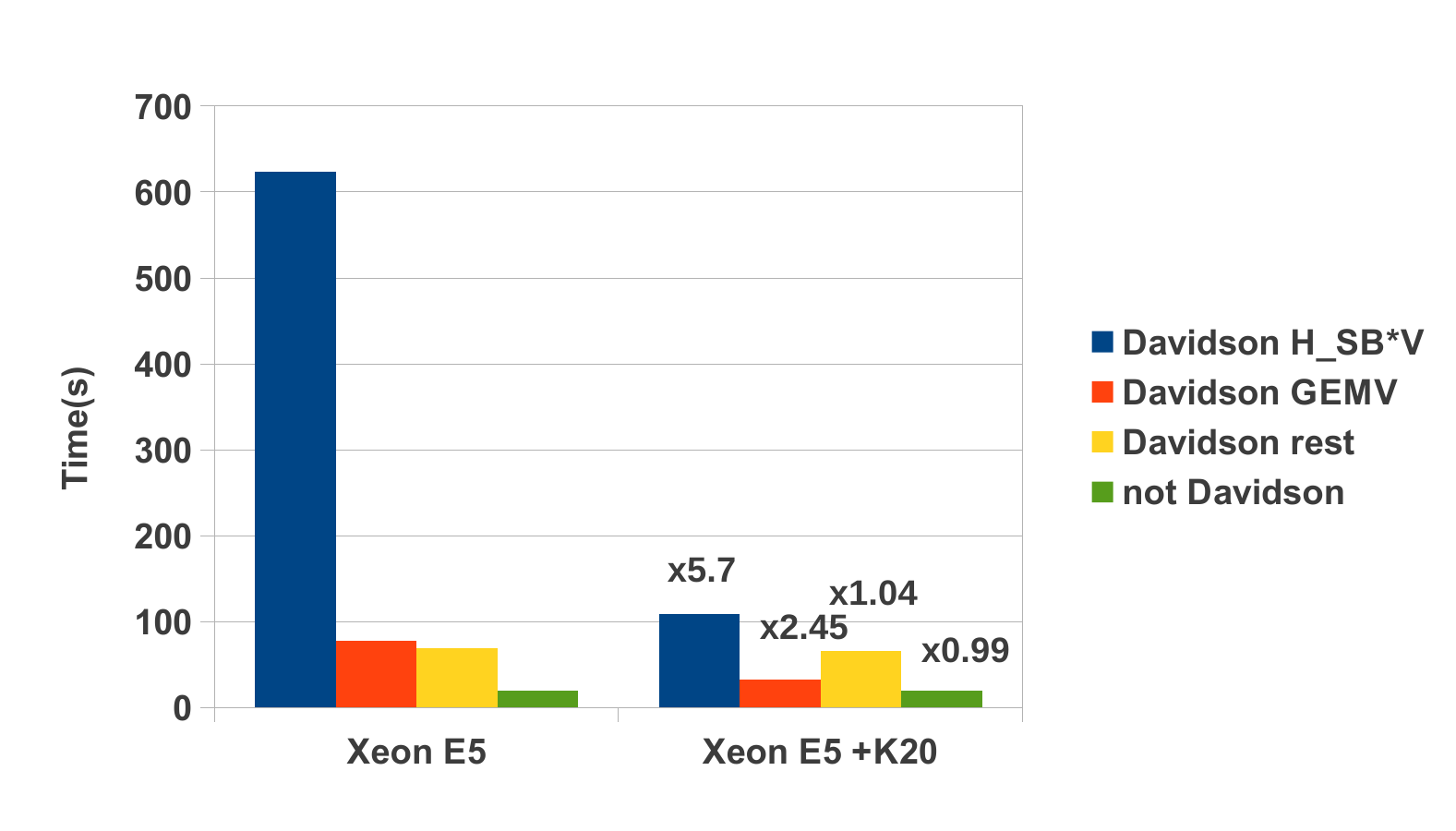}
  \end{center}
  \caption{K20, Heisenberg model: Acceleration of different parts of the algorithm is compared for $m=4096$.}
  \label{FigInnerHeis}
\end{figure}

\begin{figure}[h!]
  \begin{center}
    \includegraphics[height=5cm]{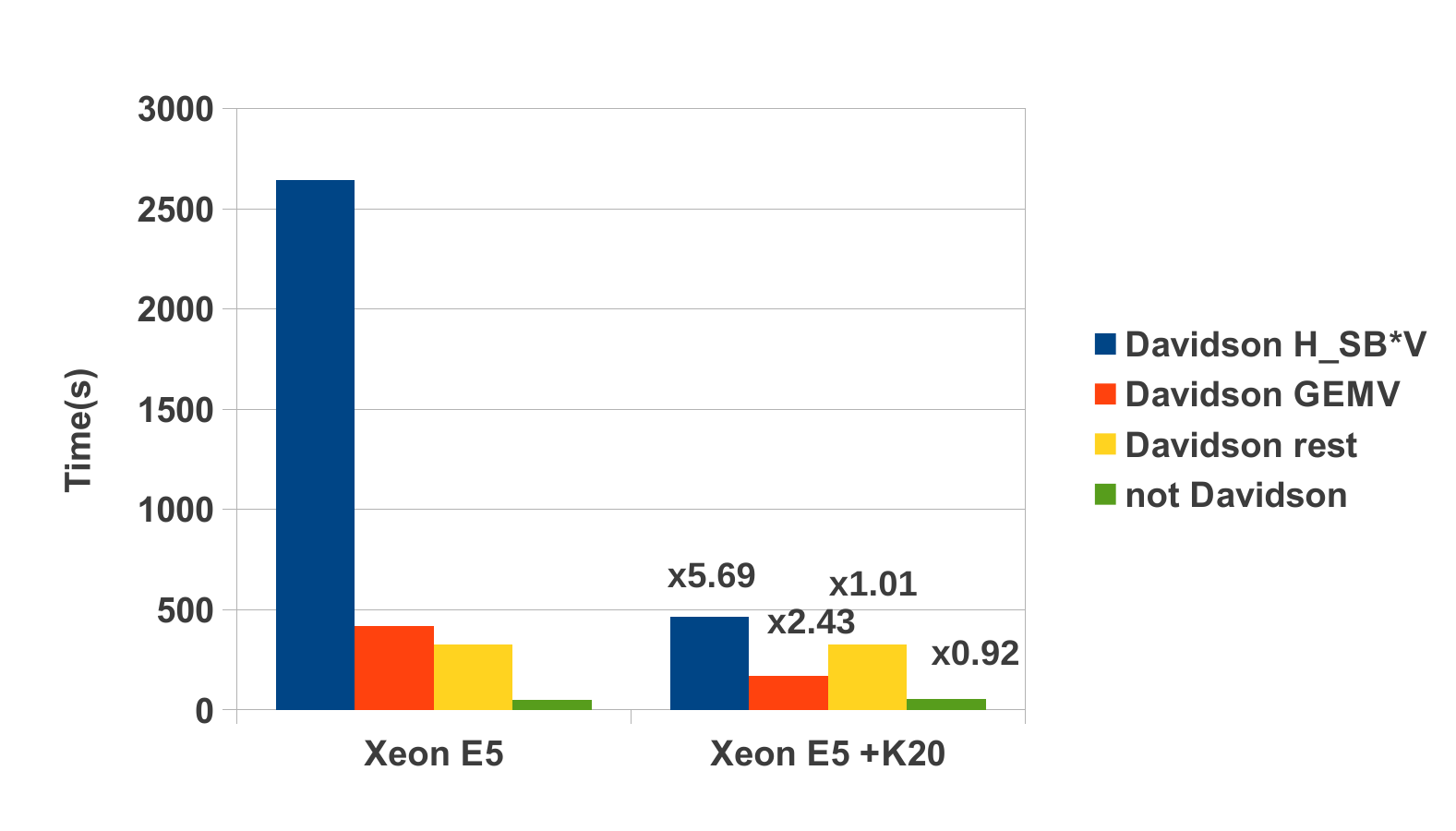}
  \end{center}
  \caption{K20, Hubbard model: Acceleration of different parts of the algorithm is compared for $m=4096$.}
  \label{FigInnerHub}
\end{figure}

In case of K20 the acceleration of the projection and the matrix-vector operations is compared in Figures~\ref{FigInnerHeis} and~\ref{FigInnerHub}. 
The projection is accelerated by $5.7$ times which is in accordance with the theoretical performance capabilities of the two architectures.
Currently on Xeon processor (see Figure~\ref{FigPercentage}) the projection operation is only accounted for $75\%$ of the total run-time, therefore, the overall acceleration is also affected by the rest of the operations of the Davidson algorithm.
Fortunately, as the number of retained states ($m$) increases the time-dominance of the projection also increases, which anticipates even better acceleration for real-world simulations with large~$m$.

As the acceleration of the full Davidson algorithm can be limited by the GPU memory, an adaptive solution shall be implemented which accelerates as much of the algorithm as possible.
Currently four matrix-vector operation of the algorithm is accelerated in case of sufficient GPU memory, however, latter acceleration of the rest of the operations will be implemented as well.


\section{Conclusion}\label{SecCon}

In this paper acceleration of the DMRG algorithm using novel kilo-processor architectures (GPU, FPGA) has been investigated.
The GPU architecture has been found to a promising accelerator, as the most time-dominant step of the algorithm, the projection operation, can be formulated as independent dense matrix multiplications, which are ideal workload for GPUs.
Moreover, in case of high-end GPUs the acceleration of the projection is so remarkable, that it is worth to consider the acceleration of the rest of the algorithm to obtain a decent overall speed-up.
In the presented implementation some asymmetric matrix-vector multiplications of the diagonalization have been identified as the second most time-dominant part of the algorithm and a new CUDA kernel has been designed to efficiently accelerate these operations.
The resulting parallelized DMRG implementation is a hybrid CPU-GPU solution, which distributes the workload according to the performance and memory capabilities of the configurations and anticipates a straightforward multi-GPU extension, which is part of our current developments.

In subsequent works our DMRG implementation will be utilized to investigate more complex models focusing on physical problems of current interests, which are hard to treat due to their high computational demands.
For example, high-spin fermionic systems relevant for ultracold atomic experiments and extensions to treat ab-initio quantum chemical applications~\cite{Schneider2013,Reiher2010,Chan2012} which are already under progress.
Furthermore, a straightforward generalization of the presented algorithm to accelerate tensor network states (TNS) algorithms~\cite{Orus2013} is another promising research direction.

In general, FPGA chips have lower operating frequencies than in case of GPU architectures and the attached on-board memories are also smaller and slower.
FPGAs can outperform other architectures in such problems where a custom arithmetic unit can reach a significantly better utilization, however, in case of dense matrix operations of the DMRG algorithm nearly ideal utilization of the CUDA cores is reached.
As the estimated performance of the considered high-end FPGAs is around the mid-range GPUs and the development time for FPGA is significantly longer, the GPU architecture is preferred for the acceleration of the DMRG algorithm.

\section*{Acknowledgement}
This research was supported by TÁMOP-4.2.2.C-11{\hskip 0pt}/1/KONV-2012-0004, TÁMOP-4.2.1./B-11/2/KMR-2011-002, TÁMOP-4.2.2./B-10/1-2010-0014 and the Hungarian Research Fund (OTKA) under Grant Nos. NN110360, K100908 and K84267.

\bibliographystyle{elsarticle-num}
\bibliography{IEEEexample}
\end{document}